\documentclass[reprint,preprintnumbers,amsmath,amssymb]{revtex4}
\usepackage{amssymb,graphicx}
\usepackage{amsmath,bm}

\usepackage{amssymb,graphicx}

\newcommand{\bea}{\begin{eqnarray}}
\newcommand{\eea}{\end{eqnarray}}
\newcommand{\bes}{\begin{subequations}}
\newcommand{\ees}{\end{subequations}}

\begin{document}
\title{Mixed solitons in  ($2+1$) dimensional multicomponent long-wave$-$short-wave system}
\author{T. Kanna\footnote{e-mail: kanna{\_}phy@bhc.edu.in}}
\affiliation{Post Graduate and Research Department of Physics, Bishop Heber College, Tiruchirapalli--620 017, India.}
\author{M. Vijayajayanthi\footnote{Corresponding author e-mail: vijayajayanthi.cnld@gmail.com}}
\affiliation{Department of Physics, Anna University, Chennai--600 025, India.}
\author{M. Lakshmanan\footnote{e-mail: lakshman@cnld.bdu.ac.in}}
\affiliation{Centre for Nonlinear Dynamics, School of Physics, Bharathidasan University, Tiruchirapalli--620 024, India.}

\date{\today}

\begin{abstract}
We derive a (2+1)-dimensional multicomponent long-wave$-$short-wave resonance interaction (LSRI) system as the evolution equation for propagation of $N$-dispersive waves in weak Kerr type nonlinear medium in the small amplitude limit. The mixed (bright-dark) type soliton solutions of a particular (2+1)-dimensional multicomponent LSRI system, deduced from the general multicomponent higher dimensional LSRI system, are obtained by applying the Hirota's bilinearization method. Particularly, we show that the solitons in the LSRI system with two short-wave components behave like scalar solitons. We point out that for $N$-component LSRI system with $N>3$, if the bright solitons appear in atleast two components,  interesting collision behaviour takes place resulting in energy exchange among the bright solitons. However the dark solitons undergo standard elastic collision accompanied by a position-shift and a phase-shift. Our analysis on the mixed bound solitons shows that the additional degree of freedom which arises due to the higher dimensional nature of the system results in a wide range of parameters for which the soliton collision can take place.
\end{abstract}

\maketitle
\thispagestyle{empty}
\section{Introduction}\label{intro}
The nonlinear interaction of multiple waves results in several new physical processes \cite{{kiv},{akm},{scot}}. It has been shown in the two layer fluid model that resonance between the long-wave component and short-wave component occurs when the phase velocity of the former matches with the group velocity of the latter \cite{oikawa}. This is a ubiquitous phenomenon which appears in hydrodynamics \cite{grim}, bio-physics \cite{boiti}, plasma physics \cite{zakh}, and in nonlinear optical systems \cite{kivol}. Though there are many  studies on the long-wave$-$short-wave resonance interaction (LSRI) in one dimension \cite{{zakh},{kivol},{yaji76},{ma78},{ma79},{funa}}, results are scarce for multicomponent higher dimensional LSRI system. In the context of nonlinear optics, the interaction of bright and small amplitude dark pulses in optical fiber is governed by the integrable Zakharov model \cite{kivol, zakh}.

The resonance interaction of long-wave with short-wave  was first investigated by Benney for capillary-gravity waves in deep water \cite{benny}. In this case simple interaction equations cannot be obtained, because for deep water waves there is no wave in the long wavelength limit. However, simple interaction equations can be deduced in a stable stratified fluid for oblique propagation of long and short-waves \cite{grim}. The single component two-dimensional LSRI equation for a two-layer fluid model has been derived in Ref. \cite{oikawa} by using the multiple scale perturbation method and bright and dark type one- and two-soliton solutions have been reported. In Ref. \cite{ohta7}, Ohta {\it et al} derived an integrable two component analogue of the two-dimensional LSRI system as a governing equation for the interaction of the nonlinear dispersive waves by applying the reductive perturbation method. It is worth noting that there exists several articles \cite{oc1,oc2,rev} on this perturbation approach which is based on a consistent and mathematically rigorous expansion of the linear dispersion relation including the nonlinear optical response of the medium.  It leads to a new equation for self-focusing of extremely focused short-duration intense pulses \cite{oc1} and also to a general propagation equation for the pulse envelope of an electromagnetic field in an isotropic nonlinear dispersive medium \cite{oc2} with all orders of dispersion, diffraction and nonlinearity.  Very recently, the non-integrable three component Gross-Pitaevskii equations have been reduced to single component Yajima-Oikawa system by using multiple scale method \cite{nista}. In another recent work \cite{visi}, the one-dimensional integrable two-component Zakharov-Yajima-Oikawa equation has been derived using multiple scale method and special bright-dark one-soliton solutions have been reported.

In Ref. \cite{tklsri}, we have obtained general bright $M$-soliton solution, for arbitrary $M$, of the same integrable multicomponent ($2+1$)D-LSRI system (see Eq. (\ref{model})) considered in the present paper. And also, the bright soliton bound states of the same system have been analyzed in Ref. \cite{epjsk}.  Our earlier work \cite{tklsri} on bright soliton solutions of the multicomponent LSRI system shows that the role of long interfacial wave is to induce nonlinear interaction among the short-wave components resulting in non-trivial (shape-changing) collision behaviour characterized by energy exchange among the two short-wave components. As far as we know, for the first time in Ref. \cite{tklsri} we have identified the shape-changing/energy sharing collisions of bright solitons in a two dimensional integrable nonlinear system. This system will act as a potential candidate for realizing soliton collision-based computing and multi-state logic \cite{jak,ste,tkpre3}. Now it is of interest to derive the general two dimensional $N$-component equations describing the interaction of several short-wave packets with long-waves in a physical set up and to look for other types of multicomponent soliton solutions.

In recent years, much attention has been paid to investigate mixed (bright-dark) soliton dynamics in different dynamical systems including nonlinear optical systems and Bose-Einstein condensates \cite{kiv,kivol,tkpra8,ohta7,shep,ohtadark,nista,visi} of coupled bright-dark solitons and to analyse their propagation properties and collision dynamics. In the present work, we derive the (2+1)-dimensional $N$-component LSRI system governing the evolution of $m$ short-waves and $n$ long-waves (with $m+n=N$) in a nonlinear dispersive medium and reduce the system to an integrable system for a particular choice of the system parameters. Then by applying the elegant Hirota's direct method to the integrable multicomponent LSRI system for a particular choice, we obtain the coupled bright-dark one- and
two- soliton solutions. We will show that  bright and dark parts of the mixed solitons in the two short-wave components case behave like scalar solitons whereas if we go for three or more short-wave components the multicomponent nature of the solitons will come into picture and one can observe interesting propagation and collision properties. It is straight-forward albeit lengthy procedure to extend the analysis to construct $M$-soliton solution, with arbitrary $M$.

The present paper is organized as below. The general $N$-component LSRI system is derived by applying the multiple scale perturbation method in the next section. In section \ref{bilinear}, bilinear equations for the integrable (2+1)-dimensional multicomponent LSRI system are given. Sections \ref{onesol} and \ref{twosol} deal with the mixed one- and two-soliton solutions of the multicomponent LSRI system. The collision dynamics of the solitons are discussed in section \ref{collision}. Discussion on the mixed soliton bound states is presented in section \ref{secbs}. The final section is allotted for conclusion.

\section{The Model}\label{secmodel}
To start with  we obtain the general two-dimensional multicomponent evolution equation for the propagation of $N$-dispersive waves in a Kerr type nonlinear medium (ex.: optical fiber, photo-refractive medium) by generalizing the approach developed  in Refs. \cite{{kivol},{ohta7}} for the two and three components case. The waves are
assumed to obey the following weakly nonlinear dispersion relations
\bea
\omega_j=\omega_j(\mathbb{K}_j;\mathbb{L}_j:|A_1|^2,|A_2|^2,...,|A_N|^2), \quad
j=1,2,3,...,N,
\label{meq1}
\eea
where $\mathbb{K}_j$ and $\mathbb{L}_j$ are the $x$ and $y$ components of the wave vector, $A_j(\equiv A_j(x,y,t))$ and $\omega_j$ are the complex amplitude and angular frequency of the $j$-th wave. Especially, in the physical setting of propagation of an incoherent self-trapped beam in a slow Kerr-like medium, the nonlinearity arising from the change in refractive index profile (say $\delta n$) created by all incoherent components of the light beam can be expressed as $\delta n=\sum_{m=1}^{M}\alpha_m |A_m|^2$, where $|A_m|^2$ is the intensity of the $m$-th incoherent component, $\alpha_m$ is the nonlinearity coefficient of the $m$-th component and $M$ denotes the total number of components. This shows that we can very well have nonlinearities which are purely dependent only on intensities even for multicomponent systems. Such media will be appropriate to realize the type of dispersion relation considered in this paper.  This type of system is known as incoherently coupled system in the context of nonlinear optics \cite{kiv}.  The fundamental carrier wave is of the form $e^{i(\mathbb{K}_0x+\mathbb{L}_0y-\omega_0t)}$. The most convenient way to derive the evolution equation for the amplitudes $A_j$'s is to Taylor expand the angular frequencies $\omega_j$'s around the $x$ and $y$ components of the wave vector of the carrier wave $\mathbb{K}_0$ and $\mathbb{L}_0$, respectively, and the central frequency $\omega_0$ at $|A_j|=0$, as below:
\bea
 (\omega_j-\omega_0)&=&(\omega_{j,\mathbb{K}_j})_0 \Delta \mathbb{K}_j +(\omega_{j,\mathbb{L}_j})_0 \Delta \mathbb{L}_j + \frac{1}{2} (\omega_{j,\mathbb{K}_j\mathbb{K}_j})_0 \Delta \mathbb{K}_j^2 + \frac{1}{2} (\omega_{j,\mathbb{L}_j \mathbb{L}_j})_0 \Delta \mathbb{L}_j^2 ~~~~~~~\nonumber\\
  &&+ (\omega_{j,\mathbb{K}_j \mathbb{L}_j})_0 (\Delta \mathbb{K}_j )(\Delta \mathbb{L}_j) +  \sum_{m=1}^N \left( \omega_{j, |A_m|^2} \right)_0|A_m|^2+..., \quad j=1,2,...,N,~~~~
\label{meq2}
\eea
where $\Delta \mathbb{K}_j=\mathbb{K}_j-\mathbb{K}_0$, $\Delta \mathbb{L}_j=\mathbb{L}_j-\mathbb{L}_0,~ j=1,2,...,N$. In this Taylor expansion and in the following, the subscript `$0$' given in Eq. (\ref{meq2}) as $\left (~~\right)_0$ represents the fact that the quantity appearing inside the bracket is evaluated at $\mathbb{K}_j=\mathbb{K}_0,~\mathbb{L}_j=\mathbb{L}_0,~\omega_j=\omega_0$ and $|A_j|=0$. In Eq. (\ref{meq2}), $\omega_{j,\mathbb{K}_j}=\frac{\partial\omega_j}{\partial \mathbb{K}_j}$, $\omega_{j,\mathbb{L}_j}=\frac{\partial\omega_j}{\partial \mathbb{L}_j}$, $\omega_{j,\mathbb{K}_j\mathbb{K}_j}=\frac{\partial^2\omega_j}{\partial \mathbb{K}_j^2}$, $\omega_{j,\mathbb{L}_j\mathbb{L}_j}=\frac{\partial^2\omega_j}{\partial \mathbb{L}_j^2}$, $\omega_{j,\mathbb{K}_j \mathbb{L}_j}=\frac{\partial^2\omega_j}{\partial \mathbb{K}_j \partial \mathbb{L}_j}$ and $\omega_{j, |A_m|^2}=\frac{\partial \omega_j}{\partial |A_m|^2}$. Then by replacing $(\omega_j-\omega_0)$, $\Delta \mathbb{K}_j$ and $\Delta \mathbb{L}_j$ by the operators $-i\frac{\partial}{\partial t}$,  $i\frac{\partial}{\partial x}$ and $i\frac{\partial}{\partial y}$, respectively and transforming to the moving co-ordinates $x'=x-\overline{\omega}_1 t$, $y'=y-\overline{\omega}_2 t$, $t'= t$, with the assumption that beyond a particular component (say $q^{{th}}$) all the derivatives $\left(\frac{\partial \omega_j}{\partial \mathbb{K}_j}\right)_0$, $j=q+1,q+2,...,N$, are same and so also the derivatives $\left(\frac{\partial \omega_j}{\partial \mathbb{L}_j}\right)_0$, i.e., $\left(\omega_{q+1,\mathbb{K}_{q+1}}\right)_0 \left(\equiv \frac{\partial \omega_{q+1}}{\partial \mathbb{K}_{q+1}}\right) = \left(\omega_{q+2,\mathbb{K}_{q+2}}\right)_0=...=\left(\omega_{N,\mathbb{K}_{N}}\right)_0 \equiv \overline{\omega_1}$ (say) and $\left(\omega_{q+1,\mathbb{L}_{q+1}}\right)_0 = \left(\omega_{q+2,\mathbb{L}_{q+2}}\right)_0=...=\left(\omega_{N,\mathbb{L}_N}\right)_0 \equiv \overline{\omega_2}$ (say), and omitting the primes for simplicity of notation we get
\bes\bea
 i A_{j,t}&+&i v_{jx} A_{j,x}+i v_{jy} A_{j,y} + C_1^{(j)} A_{j,xx}+ C_2^{(j)} A_{j,yy}+ C_3^{(j)} A_{j,xy} + \sum_{l=1}^N B_l^{(j)}
|A_l|^2 A_j =0,~\label{4a}\\
  i A_{p,t} &+& C_4 A_{p,xx}+ C_5 A_{p,yy}+ C_6 A_{p,xy} + \sum_{l=1}^N B_l^{(p)} |A_l|^2 A_p =0, \quad  \\
&&~~~~~~~~~~~~~~~~~~~~~~~~~~~~~~~~~~~~~~~~~~~~~~~~~~j=1,2,...,q,\quad
p=q+1,q+2,...,N.\nonumber
\eea\label{meq4}\ees
Here the independent variables appearing in the suffixes after the comma denote partial derivatives with respect to that variables and the  group velocities of the $j^{th}$ component along the $x$ and $y$ directions are $v_{jx}=((\omega_{j,\mathbb{K}_j})_0-{\overline{\omega}_1})$ and $v_{jy}=((\omega_{j,\mathbb{L}_j})_0-{\overline{\omega}_2})$, respectively. Various quantities in the above equations (\ref{meq4}) are defined as $C_1^{(j)}= \left(-\frac{\omega_{j,\mathbb{K}_j\mathbb{K}_j}}{2} \right)_0$, $C_2^{(j)}= \left(-\frac{\omega_{j,\mathbb{L}_j \mathbb{L}_j}}{2} \right)_0$, $C_3^{(j)}= \left(-{\omega_{j,\mathbb{K}_j \mathbb{L}_j}} \right)_0$, $C_4=\left(\frac{-\overline{\omega}_{1,\mathbb{K}_N}}{2}\right)_0$, $C_5=\left(\frac{-\overline{\omega}_{2,\mathbb{L}_N}}{2}\right)_0$, $C_6=\left(-\overline{\omega}_{1,\mathbb{L}_N}\right)_0$, $B_j^{(i)}=\left(\frac{\partial \omega_i}{\partial |A_j|^2}\right)_0 \equiv \left(\omega_{i,|A_j|^2}\right)_0$, $i,j=1,2,3,...,N$.

We consider the case where the first $q$-components are in the anomalous dispersion region and the remaining ($N-q$)-components are in the normal dispersion regime. Following Ref. \cite{kivol}, the solutions of (\ref{meq4}) are sought in the form
\bes\bea
&& A_j= \psi_j(x,y,t) e^{i \delta_j t}, \quad j=1,2,...,q,\\
&& A_p= (u_0+a_p(x,y,t)) e^{i[\Lambda_p t+\vartheta_p (x,t)]}, \quad
p=q+1, q+2,..., N, \eea \label{meq5}\ees
where $\delta_j= \displaystyle\sum_{l=q+1}^N B_l^{(j)} u_0^2$, $\Lambda_p=\displaystyle\sum_{l=q+1}^N B_l^{(p)} u_0^2$ and $a_p$'s are assumed to take only small values.

Substituting equations (\ref{meq5}) in (\ref{meq4}a) and neglecting the higher order terms involving $(a_p,\vartheta_p)$ and also their derivatives result in the equation
\bea
&& i(\psi_{j,t} + v_{jx}\psi_{j,x} + v_{jy} \psi_{j,y}) + C_1^{(j)} \psi_{j,xx}+C_2^{(j)} \psi_{j,yy}+C_3^{(j)} \psi_{j,xy} \nonumber\\
 && \quad + \left(\sum_{l=1}^q B_l^{(j)} |\psi_l|^2\right) \psi_j +\left(\sum_{p=q+1}^N B_p^{(j)} (2u_0 a_p)\right) \psi_j=0, \quad j=1,2,...,q. \label{meq6a}
\eea
In a similar manner, by incorporating (\ref{meq5}) in (\ref{meq4}b) and collecting the real and imaginary parts, we arrive at a set of two  coupled equations. The resulting coupled equations can be grouped together to obtain the following equation by differentiating the imaginary part equation twice with respect to `$t$' and making use of the real part equation:
\bea
 a_{p,tt} &+& C_4^2 a_{p,xxxx}+C_4C_5 a_{p,xxyy}+ C_4C_6 a_{p,xxxy} + C_4u_0\left(\sum_{j=1}^q B_j^{(p)} |\psi_j|_{xx}^2\right)  \nonumber\\
&&  + 2C_4u_0^2 \sum_{l=q+1}^N B_l^{(p)} a_{l,xx} =0 , \quad p=q+1,q+2,...,N.~~~~~~~ \label{meq9}
\eea
Equations (\ref{meq6a}) and (\ref{meq9}) are general equations describing the two dimensional propagation of $q$ waves in the anomalous dispersion region and $(N-q)$-waves in the normal dispersion region. For $N=3$ case with $q=2$ the system has two short-wave components and one long-wave component and coincides with the corresponding equations presented in Ref. \cite{ohta7}. One can notice from the general form of equations (\ref{meq6a}) and (\ref{meq9}) that for the same $N(=3)$ but with different $q$ value, (say $q=1$), there is another possibility which will have one short-wave component and two long-wave components, and ultimately the dynamics will be different from the $q=2$ case. Also, this systematic generalization to the $N$ component case is necessary to identify the way by which the additional wave components (modes) in the normal dispersion regime alter the governing equation for the three components case given in Ref. \cite{ohta7}.

To deduce an integrable equation associated with the combined systems (\ref{meq6a}) and (\ref{meq9}) we choose all the $ B_l^{(p)}$'s, $l,p=q+1,...,N,$ in Eq. (\ref{meq9}) to be equal to a constant value (say $-\gamma_1$, $\gamma_1>0$). Then we get
\bea
a_{p,tt}&+& C_4^2 a_{p,xxxx}+C_4C_5 a_{p,xxyy}+ C_4C_6 a_{p,xxxy} + C_4u_0\left(\sum_{j=1}^q B_j^{(p)} |\psi_j|_{xx}^2\right)  \nonumber\\
&& - 2C_4 \gamma_1 u_0^2 \sum_{l=q+1}^N a_{l,xx}=0, \quad p=q+1,q+2,...,N.~~~~~~~ \label{meq9a}
\eea

In the following, we investigate the cumulative effect of the small amplitudes $a_p$'s on the short-wave components by considering the superposition  of these amplitudes involving only the sum of all the amplitudes and neglect all other combinations as they will be small due to the smallness of $a_p$'s. Particularly, we add all the $a$-equations and define $\sum_{l=q+1}^N a_l=L$. By doing so we get
\bea
&&L_{tt}+C_4^2L_{xxxx}+C_4C_5L_{xxyy}+C_4C_6L_{xxxy}+C_4 u_0\left(\sum_{p=q+1}^N \sum_{j=1}^q B_j^{(p)} |\psi_j|_{xx}^2\right) \nonumber\\
&& \qquad -2C_4[N-q] u_0^2\gamma_1 L_{xx} = 0. \label{meq9b}
\eea
The dispersion relation for the linear excitation corresponding to the long-wave components is found as
\bea
\Omega^2=c^2k^2 \left[1+\frac{C_4^2}{c^2}{k^2}+\frac{C_4C_5}{c^2}l^2+\frac{C_4C_6}{c^2}kl
\right], \label{meq10a} \eea
where $c^2=2C_4 u_0^2(N-q)\gamma_1$. Note that the velocity of the linear excitation depends upon the number of components $N$ and increases as we increase the number of components (modes) in the normal dispersion regime. Thus our systematic generalization to $N$-component case shows that by altering the number of components in the normal dispersion region one can change the velocity of the pulse.

Next we apply the multiple scale approximation method to derive the two-dimensional multicomponent LSRI system as in Ref. \cite{ohta7}. We re-scale the variables $t$, $x$, $y$, $a_p$ and $\psi_j$ as
\bea t''=\epsilon t, \quad x''=\sqrt{\epsilon} (x+ct), \quad
y''=\epsilon y, \quad
a_p={\epsilon} ~\hat{a}_p ,\;\;
\psi_j=\epsilon^{3/4} S^{(j)},
\label{meq10}\eea
where $c$ is as defined above after the dispersion relation (\ref{meq10a}). Then the following set of equations results from Eq. (\ref{meq9b}) at the order $\epsilon^{5/2}$
\bes\bea
2c L_{xt}+ C_4 u_0 \left(\sum_{p=q+1}^N \sum_{j=1}^q B_j^{(p)} |S^{(j)}|_{xx}^2\right)=0, \quad
p=q+1,q+2,...,N.~~ \label{meq11}
\eea

At the order of $\epsilon^{5/4}$, we notice from Eq. (\ref{meq6a}) that all the group velocities of the short-wave components along the $x$ direction are the same and their magnitudes are equal to the phase velocity of the long-wave component `$c$' (i.e. $v_{jx}=-c,~j=1,2,...,q$). This is the condition for resonant interaction between long-waves and short-waves. Equation (\ref{meq6a}) reduces to the following set of coupled equations at the order of $\epsilon^{7/4}$ after replacing $v_{jx}$ by $c$ and rescaling of the variables as defined in equation(\ref{meq10}),
\bea
&& i\left(S_t^{(j)} + v_{jy} S_y^{(j)}\right) +C_1^{(j)} S_{xx}^{(j)} + \left(2u_0 \sum_{p=q+1}^N B_p^{(j)} \hat{a}_p\right) S^{(j)}=0, \quad j=1,2,...,q.
\label{meq12a}\eea \label{meq12}\ees
In Eq. (\ref{meq12}), after applying the transformations (\ref{meq10}) the double primes in the new variables `$t$', `$x$' and `$y$' are dropped, for convenience.

Equation (\ref{meq12}) is the multicomponent LSRI system in ($2+1$)-dimensions which is non-integrable in general. By suitably choosing the constants $B_j^{(p)}$'s, $B_p^{(j)}$'s, $C_4,\; C_1^{(j)}$, $j=1,2,...,q$, $p=q+1,...,N$, and $\gamma_1$, along with the assumption that there is no group velocity delay between the short-wave components, we arrive at the following ($q+1$)-component (2+1)-dimensional LSRI system for the $2$-dimensional propagation of $N$ dispersive waves in weak Kerr-like nonlinear media,
\bes \bea
&& i\left(S_t^{(j)}+S_y^{(j)}\right) - S_{xx}^{(j)} +L S^{(j)} = 0, \quad j=1,2,...,q,\\
&& L_t = 2 \sum_{j=1}^q |S^{(j)}|^2_x.
\eea \label{model} \ees
In Eq. (\ref{model}), the subscripts denote partial derivatives with respect to those independent variables. As mentioned in the introduction, we have obtained more general bright $M$-soliton solution, with arbitrary $M$, of Eq. (\ref{model}) \cite{tklsri}. In reference \cite{tklsri}, we have expressed the bright $M$-soliton solution of (\ref{model}) in Gram determinant form and explicitly proved that the general multisoliton solution indeed satisfies the bilinear equations. We have also   pointed out in the same work that for the two short-wave components case ($q=2$), the bright soliton solutions reported by Ohta {\it et al} in Ref. \cite{ohta7} follow as special cases of our general multi-soliton solutions \cite{tklsri}. As the $(q+1)$-component LSRI system (\ref{model}) admits $M$-soliton solution, for arbitrary $M$ \cite{tklsri}, the system can be integrable \cite{{hirotabook},{hiet}}. The study on the other integrability aspects of Eq. (\ref{model}) is under progress and will be published elsewhere.

\section{Hirota's Bilinearization Method for the ($2+1)D$ Multicomponent LSRI system}\label{bilinear}
There are several efficient analytical tools to construct various types of localized structures for nonlinear evolution equations, which include inverse scattering transform method, Hirota's bilinearization method, Darboux transformation method, Lie symmetry analysis, tanh method, etc.  By performing the bilinearizing transformations using Hirota's direct method \cite{hirotabook,bull}, we construct soliton solutions of Eq. (12) in this paper.   In Ref. \cite{gramm,jh}, an extension of Hirota's bilinear formalism (i.e. multilinear operator) that can encompass any degree of multilinearity has been presented.  Using this generalization of Hirota's method, propagation of a monochromatic laser beam coupled to its second and third harmonics in a nonlinear medium has been studied by  V. Cao Long et al. \cite{long1,long2}.  Recently, a bilinearization procedure with a set of generalized bilinear differential operators different from the standard Hirota's operators, having nice mathematical properties has been proposed \cite{ma1}. Apart from this, in Ref. \cite{ma2}, it has been pointed out that by employing Lie symmetry approach to the one-dimensional scalar nonlinear Schr\"odinger equation and  by performing a direct search  various exact new interesting solutions can be obtained. The Lie algebraic structure of system (\ref{model}), specifically for $q=1$ has been discussed in Ref. \cite{lie}. Indeed, it will be an interesting future direction to compute the Lie symmetries of the multicomponent system (\ref{model}) for $q>2$. In this connection, we may also mention that in the past Lie symmetries of certain (2+1) dimensional systems have been constructed by first finding the symmetries of a given (2+1) dimensional system and then reducing it to a (1+1) dimensional system, which on identifying its own Lie symmetries can be reduced to ordinary differential equations. In certain cases, the (2+1) dimensional evolution equations also lead to the identification of infinite dimensional Kac-Moody-Virasoro algebras \cite{symmetry}. Apart from the above one can also construct the various interaction solutions of the present system using the Wronskian technique as done in Ref. \cite{wronskian} for the KdV equation. The multicomponent system (\ref{model}) will admit a richer solution structure that may comprise bright solitons, bright-dark solitons, dark solitons, dromions, rational solutions, periodic solutions, elliptic function solutions, and so on.

The present work specifically deals with the study of interesting bright-dark (mixed) solitons  of Eq. (\ref{model}), comprising of  $m$ bright parts and $n$ dark parts, such that $m+n=q$. These solitons are usually referred as ``symbiotic" solitons as the bright part cannot be supported in a stand-alone fashion and exists only due to the presence of its dark counterpart. These bright-dark solitons are of much theoretical and experimental interest and significant attention has been paid to investigate these intriguing vector solitons as pointed out in the introduction. In the following, we will employ the Hirota's direct method to construct such coupled bright-dark (mixed) soliton solutions for the system (\ref{model}) which can find  application in various frontier areas like nonlinear optics, water waves and Bose-Einstein condensates.

To construct the mixed type soliton solutions, we perform the bilinearizing transformations, $S^{(j)}=\frac{g^{(j)}}{f}$, $S^{(m+l)}=\frac{h^{(l)}}{f}$, $L=-2\frac{\partial^2}{\partial x^2}(\ln{f})$, $j=1, 2, \ldots, m$, and $l=1, 2, \ldots, n$, $(m+n=q)$, where $g^{(j)}$'s and $h^{(l)}$'s are arbitrary complex functions of $x, y$ and $t$ while $f$ is a real function. The resulting bilinear equations are
\bes
\bea
 &&D_1\left(g^{(j)}\cdot f\right)=0,\quad  j=1, 2, \ldots,m,\\
&& D_1\left(h^{(l)}\cdot f\right)=0,\quad l=1, 2, \ldots, n,\\
 &&D_2(f\cdot f)=-2\Bigg(\sum_{j=1}^{m}g^{(j)}g^{(j)*}+\sum_{l=1}^{n}h^{(l)}h^{(l)*}\Bigg),
\eea
\label{beq}
\ees
where $D_1=i(D_t+D_y)-D_x^2$ and $D_2=(D_t D_x-2\lambda)$, $D$-s are the standard Hirota's bilinear operators \cite{hirotabook}, $*$ stands for complex conjugation and $\lambda$ is a constant yet to be determined. One can have bright solitons for the choice $\lambda=0$ in $D_2$ \cite{{ohta7},{tklsri}} and the bright soliton collision dynamics of system (\ref{model}) has been discussed in Ref. \cite{{tklsri}}. However for non-vanishing `$\lambda$' values, the system can admit coupled bright-dark and dark-dark type soliton solutions. In this paper, we focus only on mixed (bright-dark) solitons corresponding to mixed type boundary conditions, that is, $S^{(j)}, ~L \overset{x,y,t \rightarrow \pm \infty}{\longrightarrow}~~ 0$, $S^{(m+l)} \overset{x,y,t \rightarrow \pm \infty}{\longrightarrow} \mbox{constant}$, $j=1,2,...,m, ~ l=1,2,...,n$.

This procedure can be very well applied to construct the dark-dark soliton solutions also. Here for convenience we consider the first `$m$' short-wave components to be comprised of bright parts of the mixed solitons and the remaining $n(\equiv q-m)$ components to exhibit dark parts of the mixed solitons. To construct the mixed soliton solutions we expand the variables $g^{(j)}$'s, $h^{(l)}$'s and $f$ as power series expansions in a standard way \cite{{tkpra8},{hirotabook}}. After solving the resultant set of equations recursively we can obtain the explicit forms of $g^{(j)}$'s, $h^{(l)}$'s and $f$ and hence the multisoliton solutions can be constructed.

\section{Multicomponent mixed type one-soliton solution}\label{onesol}
The mixed one-soliton solution of (\ref{model}) with $m$-bright and $n$-dark parts can be obtained by restricting the power series expansions as $g^{(j)}=\chi g_1^{(j)}$, $h^{(l)}=h_0^{(l)}(1+\chi^2 h_2^{(l)})$, $f=1+\chi^2 f_2$ and by solving the resulting equations, after their substitution into the bilinear equations (\ref{beq}) at various powers of $\chi$ recursively. The mixed one-soliton solution can be expressed in the following standard form,
\bes\bea
 && S^{(j)}=A_j k_{1R}~ \mbox{sech}\left(\eta_{1R}+\frac {R}{2}\right) e^{i\eta_{1I}}, \quad j=1,2,...,m,\\
 && S^{(m+l)}= \rho_l \;e^{i(\zeta_l+\phi_l+\pi)}\left[\mbox{cos}(\phi_l)\;\mbox{tanh} \left(\eta_{1R}+\frac {R}{2}\right) + i~\mbox{sin}(\phi_l)\right], \quad l=1,2,...,n,\\
 && L = -2 k_{1R}^2~\mbox{sech}^2 \left(\eta_{1R}+\frac{R}{2}\right).
\eea\label{solc1}
The various quantities appearing in the above equations are given below:
\bea
e^{R} &=& \frac{1}{4} \left(\sum_{j=1}^m |\alpha_1^{(j)}|^2\right) \left(\sum_{l=1}^n |\rho_l|^2 \mbox{cos}^2(\phi_l)- \omega_{1R}k_{1R} \right)^{-1},\quad A_j = \frac{\alpha_1^{(j)} e^{-\frac{R}{2}}}{(k_1+k_1^*)}, \\
\phi_l&=&\mbox{tan}^{-1}\left(\frac{k_{1I}-m_l}{k_{1R}}\right), \quad
\eta_{1R}=k_{1R}\left[x+\left(2k_{1I}-\frac{\omega_{1R}}{k_{1R}}\right)y + \left(\frac{\omega_{1R}}{k_{1R}}\right)t \right],\\
 \eta_{1I}&=& k_{1I} x- (k_{1R}^2-k_{1I}^2+\omega_{1I})y+\omega_{1I}t, \quad \zeta_l=(m_l^2-b_l)t+b_ly+m_l x.
\eea\label{solc1}\ees

In equations (\ref{solc1}) and in the following the suffixes $R$ and $I$ of a particular quantity denote the real and imaginary parts of that quantity, respectively. Also $\alpha_1^{(j)}$'s, $j=1,2,...,m$, $\rho_l$, $\omega_1(=\omega_{1R}+i\omega_{1I})$ and $k_1=k_{1R}+ik_{1I}$ are complex parameters, while $m_l$ and $b_l$, $l=1,2,...,n$, are real parameters. The above solution is non-singular for the condition $\displaystyle\sum_{l=1}^n|\rho_l|^2\mbox{cos}^2(\phi_l)>\omega_{1R}k_{1R}$. The amplitude (peak value) of the $j^{th}$ bright part of the mixed soliton is $A_j k_{1R}$ and that of $(m+l)$-th dark part of the mixed soliton is $\rho_l$. The speed of the soliton is ${\omega_{1R}}/{k_{1R}}$. It can be noticed that both parts of the soliton have the same central position ${R}/{2k_{1R}}$. But their  phases are different. In fact, the phase of the dark component has two contributions, one from the background carrier wave and the other from $\phi_l$. The quantity $\phi_l$ indeed determines the darkness of the dark soliton. It is interesting to notice that the bright and dark parts of the mixed soliton of the LSRI system with more than two short-wave components display several interesting features in contrast to the case of just two short-wave components as will be shown. To elucidate the understanding of such behaviour we present the explicit forms of one- soliton solutions for the two and three short-wave components and analyse them in the following subsections. For brevity, in the following we refer to mixed $M$-soliton solution with  $m$-bright parts and $n$-dark parts as ($m$b-$n$d) mixed $M$ soliton solution.

\subsection{Two short-wave components ($m=1$, $n=1$) case}\label{sec2c1b1d}
This case admits only a simple type of bright-dark pair in which the bright part of mixed soliton appears in the first component and the dark part of the mixed soliton in the remaining component or vice-versa. The one-soliton solution for this case can be expressed as
\bes\bea
S^{(1)}&=& \left(\sqrt{|\rho_1|^2 \mbox{cos}^2(\phi_1)- \omega_{1R} k_{1R}}\right) ~\mbox{sech}\left(\eta_{1R}+\frac {R}{2}\right) e^{i(\eta_{1I}+\theta)}, \\
S^{(2)}&=& \rho_1 \;e^{i(\zeta_1+\phi_1+\pi)}\left[\mbox{cos}(\phi_1)\;\mbox{tanh} \left(\eta_{1R}+\frac {R}{2}\right) + i~\mbox{sin}(\phi_1)\right], \\
L &=& -2 k_{1R}^2~\mbox{sech}^2 \left(\eta_{1R}+\frac{R}{2}\right),
\eea\label{sw11}\ees
where $\frac{R}{2} = \mbox{ln}\left[\frac{|\alpha_1^{(1)}|} {2\sqrt{|\rho_1|^2 \mbox{cos}^2(\phi_1)-\omega_{1R}k_{1R}}} \right]$, $\phi_1=\mbox{tan}^{-1}\left(\frac{k_{1I}-m_1}{k_{1R}}\right)$, $\theta=\mbox{tan}^{-1}\left(\frac{\alpha_{1I}^{(1)}}{\alpha_{1R}^{(1)}}\right)$, $ \zeta_1=(m_1^2-b_1)t+b_1y+m_1x$, and $\eta_{1R}$ and $\eta_{1I}$ are given in equations (\ref{solc1}e) and (\ref{solc1}f).

The amplitude  of the bright part $\left(\sqrt{|\rho_1|^2 \mbox{cos}^2(\phi_1)- \omega_{1R}k_{1R}}\right)$ is independent of the parameter $\alpha_1^{(1)}$, but it is influenced significantly by the background carrier wave ($\rho_1$).  Such a mixed soliton at $t=-3$ and $y=-1$ is depicted in Fig. 1 for the parametric choice $k_1=-3+i$, $\omega_1=1+0.7i$, $m_1=1.4$, $\rho=1-i$, $b_1=-0.2$, $\alpha_1^{(1)}=1+i$.
\begin{figure}[h]
\centering\includegraphics[width=0.95\columnwidth]{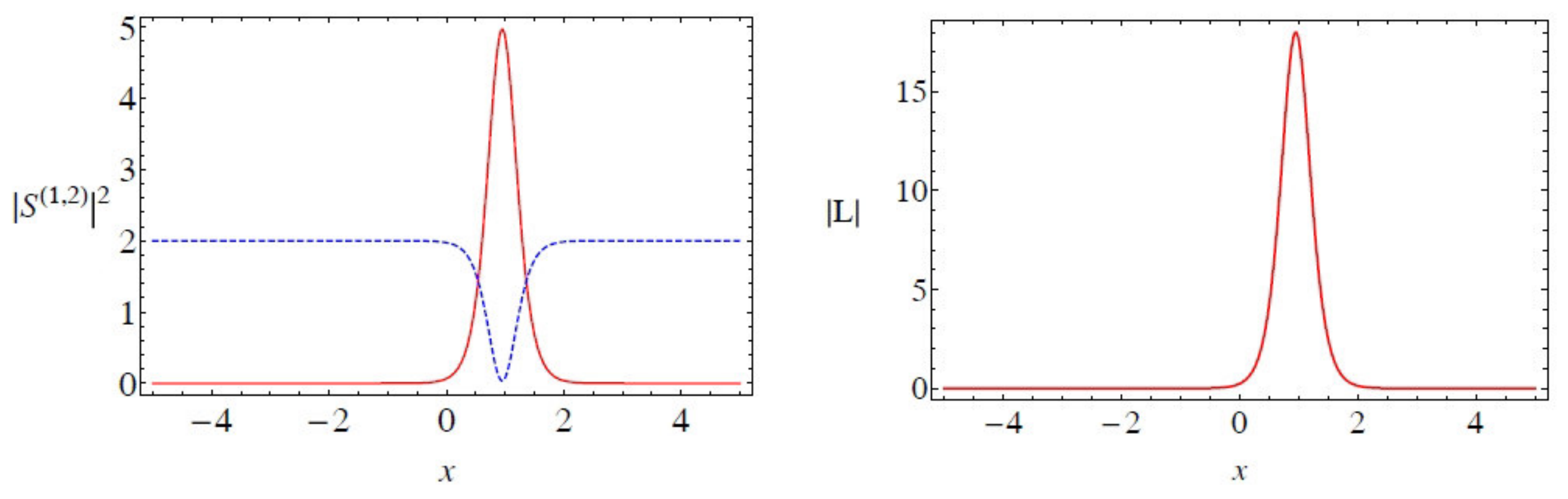}
\centering{\caption{Mixed one-soliton in  two-component LSRI system. ($|S^{(1)}|^2$ - solid curve, $|S^{(2)}|^2$ - dashed curve)}}
\label{2c1s}
\end{figure}

\subsection{Three short-wave components ($q=3$) case}
Next we consider Eq. (\ref{model}) with $q=3$. For this case the mixed soliton can be split up into bright and dark parts among the three short-wave components in two different ways. One corresponds to the ($2$b-$1$d) mixed soliton case where bright parts are in the $S^{(1)}$ and $S^{(2)}$ components while the dark part of mixed soliton appears in the $S^{(3)}$ component. The other possibility is a (1b-2d) mixed soliton case in which the bright part of the mixed soliton appears in the $S^{(1)}$ component while the dark parts are split among the remaining components $S^{(2)}$ and $S^{(3)}$.

\subsubsection{(2b-1d) mixed one-soliton solution}
\begin{figure}[h]
\centering\includegraphics[width=0.95\columnwidth]{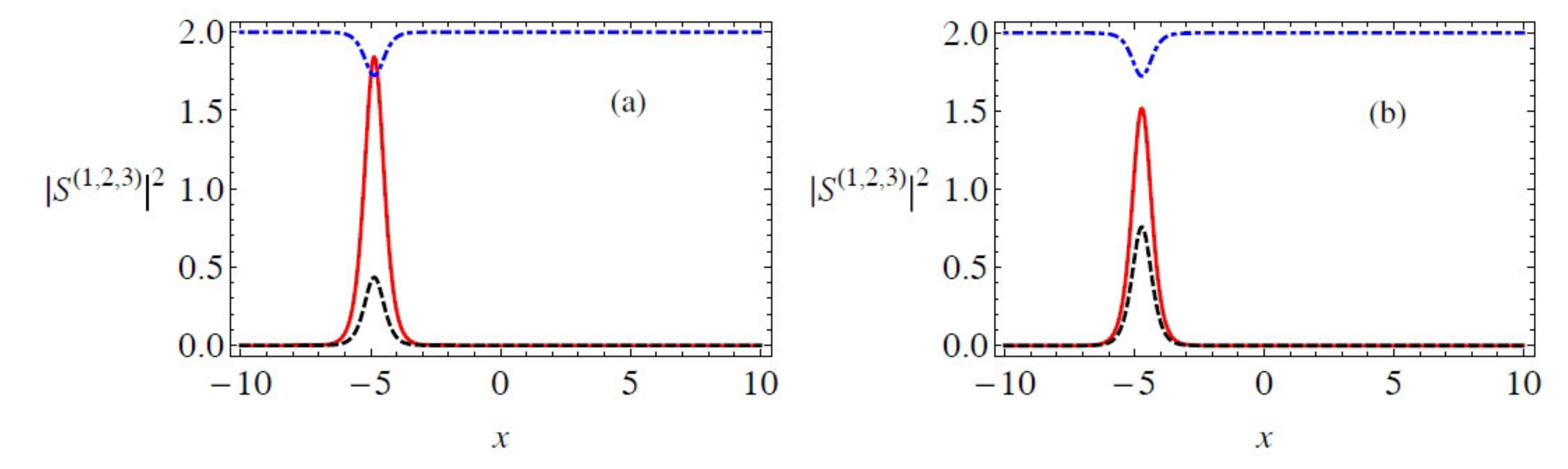}
\caption{Altering the intensity of bright soliton without affecting the dark soliton of (2b-1d) mixed one-soliton in three-component LSRI system by tuning $\alpha_1^{(j)}$ parameter. ($|S^{(1)}|^2$ - solid curve, $|S^{(2)}|^2$ - dashed curve, $|S^{(3)}|^2$ - dot-dashed curve)}
\label{3c2b1d}
\end{figure}

The one-soliton solution for this case  where the bright parts  appear in the $S^{(1)}$ and $S^{(2)}$ components while the third component  $S^{(3)}$ comprises of the dark part of the mixed soliton can be written from (\ref{solc1}) as
\bes\bea
S^{(j)}&=&A_j k_{1R}~ \mbox{sech}\left(\eta_{1R}+\frac {R}{2}\right) e^{i\eta_{1I}}, \quad j=1,2,\\
S^{(3)}&=& \rho_1 \;e^{i(\zeta_1+\phi_1+\pi)}\left[\mbox{cos}(\phi_1)\;\mbox{tanh} \left(\eta_{1R}+\frac {R}{2}\right) + i~\mbox{sin}(\phi_1)\right],
\eea\label{3c1s}\ees
where $A_j = \left(\frac{\alpha_1^{(j)}}{2k_{1R}}\right)e^{-\frac{R}{2}}$, $j=1,2$, $\zeta_1=(m_1^2-b_1)t+b_1y+m_1x$, ${R} = \ln \left[\frac{ |\alpha_1^{(1)}|^2+|\alpha_1^{(2)}|^2}{4  \left(|\rho_1|^2 \mbox{cos}^2(\phi_1)- \omega_{1R}k_{1R} \right)}\right]$, $\phi_1 = \mbox{tan}^{-1}\left(\frac{k_{1I}-m_1}{k_{1R}}\right)$, and $\eta_{1R}$ and $\eta_{1I}$ are as defined in eqn. (\ref{solc1}). $L$ takes the same form as in eqn. (\ref{sw11}c) with the above redefinition of $R$. Here one can observe that the $\alpha_1^{(j)}$-parameters appear explicitly in the amplitude of the bright soliton. The (2b-1d) one-soliton solution is characterized by twelve real parameters $\alpha_{1R}^{(1)}$, $ \alpha_{1I}^{(1)}$, $\alpha_{1R}^{(2)}$, $\alpha_{1I}^{(2)}$, $k_{1R}, k_{1I}$, $\omega_{1R}, \omega_{1I}$, $\rho_{1R}, \rho_{1I}$, $m_1$, and $b_1$ and is restricted by the condition $|\rho_1|^2\mbox{cos}^2(\phi_1) > \omega_{1R}k_{1R}$ for non-singular solutions. Such type of (2b-1d) bright one-soliton is shown in Fig. \ref{3c2b1d}(a) at $t=-3$ and $y=-1$ for the parameters $k_1=2-2i$, $\omega_1=-1-i$, $m_1=3$, $\rho_1=1-i$, $b_1=-0.2$, $\alpha_1^{(1)}=1.8+i$, and $\alpha_1^{(2)}=1$. One can also tune the intensity of bright parts without altering the depth of the dark part of the mixed soliton by suitably choosing the $\alpha_1^{(1)}$ parameter as can be evidenced from Fig. \ref{3c2b1d}(b) which is drawn for same parameter value as that of Fig. \ref{3c2b1d}(a) except for $\alpha_1^{(1)}=1+i$. The soliton appearing in the long-wave component looks similar in both the cases and so we do not present it here.

\subsubsection{(1b-2d) mixed one-soliton solution}
\begin{figure}[h]
\centering\includegraphics[width=0.955\columnwidth]{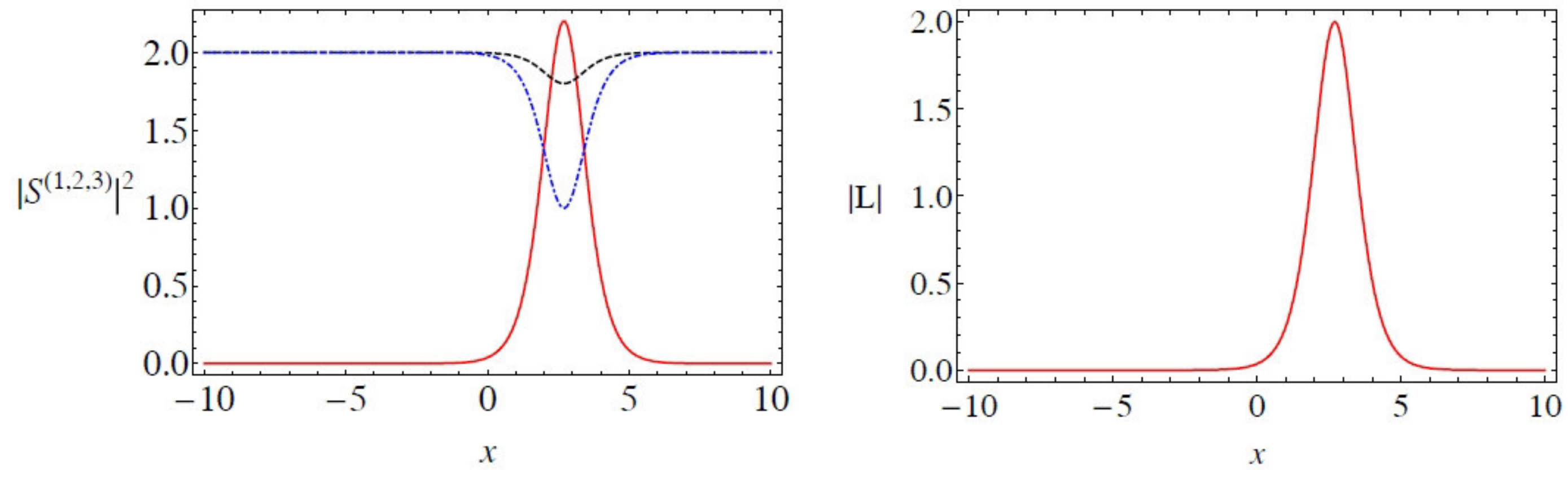}
\caption{(1b-2d) mixed one-soliton in three-component LSRI system.}
\label{3c1b2d}
\end{figure}

This case corresponds to the appearance of the bright part of the mixed soliton in the $S^{(1)}$ component while its  dark part appears in the $S^{(2)}$ and $S^{(3)}$ components. The corresponding mixed one-soliton solution is
\bes\bea
  S^{(1)}&=&\sqrt{|\rho_1|^2 \mbox{cos}^2(\phi_1) + |\rho_2|^2 \mbox{cos}^2(\phi_2) -\omega_{1R} k_{1R}} ~~\mbox{sech}\left(\eta_{1R}+\frac {R}{2}\right) e^{i(\eta_{1I}+\theta_1)}, \\
  S^{(1+l)}&=& \rho_l \;e^{i(\zeta_l+\phi_l+\pi)}\left[\mbox{cos}(\phi_l)\;\mbox{tanh} \left(\eta_{1R}+\frac {R}{2}\right) + i~ \mbox{sin}(\phi_l)\right], \quad l=1,2,\\
  L &=& -2 k_{1R}^2 ~\mbox{sech}^2 \left(\eta_{1R}+\frac {R}{2}\right),
\eea\ees
where $R = \ln \left(\frac{|\alpha_1^{(1)}|^2} {(k_1+k_1^*)^2}\right)-\ln \left(|\rho_1|^2 \mbox{cos}^2(\phi_1) + |\rho_2|^2 \mbox{cos}^2(\phi_2) -\omega_{1R} k_{1R}\right)$, $\zeta_l=(m_l^2-b_l)t+b_ly+m_lx$, $\phi_l = \mbox{tan}^{-1}\left(\frac{k_{1I}-m_l}{k_{1R}}\right)$, $l=1,2$, $\theta_1=\mbox{tan}^{-1}\left(\frac{\alpha_{1I}}{\alpha_{1R}}\right)$, and $\eta_{1R}$ and $\eta_{1I}$ are as in Eq. (\ref{solc1}). This solution is characterized by five complex parameters $\alpha_1^{(1)}$, $k_1$, $\rho_1$, $\rho_2$ and $\omega_1$ and four real parameters $b_l$ and $m_l$, $l=1,2,$ along with the condition $|\rho_1|^2 \mbox{cos}^2(\phi_1) + |\rho_2|^2 \mbox{cos}^2(\phi_2) > \omega_{1R} k_{1R}$. It can be observed from the above solution that in contrast to the (2b-1d) case, here the amplitudes of the bright and dark parts cannot be controlled by the $\alpha$ parameters. For illustrative purpose, in Fig. \ref{3c1b2d} we have shown the (1b-2d) mixed one-soliton solution for the parameters $k_1=1-i$, $\omega_1=-1-i$, $m_1=2$, $m_2=-2$, $\rho_1=1-i$, $\rho_2=1+i$, $b_1=-0.2$, $b_2=0.2$, and $\alpha_1^{(1)}=0.2-0.01i$ at $t=-3$ and $y=-1$.

\section{Multicomponent mixed type two-soliton solutions}\label{twosol}
It is a straightforward but lengthy procedure to construct the two-soliton solution. We obtain the mixed two-soliton solution of system (\ref{model}) by restricting the power series expansion for $g^{(j)}$'s, $h^{(l)}$'s and $f$ as $g^{(j)}=\chi g_1^{(j)} +\chi^3 g_3^{(j)}$, $h^{(l)}=h_0^{(l)}(1+\chi^2 h_2^{(l)} +\chi^4 h_4^{(l)})$, $f=1+\chi^2 f_2+\chi^4 f_4$, $j=1, 2, \ldots, m,~ l=1, 2, \ldots, n$, and following the standard procedure \cite{{tkpra8}}. The explicit form of ($m$b-$n$d) mixed two-soliton solution is given below.
\bes\bea
  S^{(j)}&=&\frac{1}{D}\Big(\alpha_1^{(j)}e^{\eta_1}+\alpha_2^{(j)} e^{\eta_2}+e^{\eta_1+\eta_1^{*}+\eta_2+\delta_{1j}}+e^{\eta_2+\eta_2^{*}+\eta_1+\delta_{2j}}\Big),\qquad  j=1,2, \ldots,m,~\\
  S^{(l+m)}&=&\frac{1}{D} \Big[\rho_l\;e^{i\zeta_l}\Big(1+e^{\eta_1+\eta_1^{*}+Q_{11}^{(l)}}+e^{\eta_1+\eta_2^{*}+Q_{12}^{(l)}}+e^{\eta_2+\eta_1^{*}+Q_{21}^{(l)}}\nonumber\\
  &&\qquad\qquad\qquad+e^{\eta_2+\eta_2^{*}+Q_{22}^{(l)}}+e^{\eta_1+\eta_1^{*}+\eta_2+\eta_2^{*}+Q_3^{(l)}} \Big)\Big],\qquad l=1,2, \ldots,n,\\
  L&=&-2\frac{\partial^2}{\partial x^2} \left(\ln(D)\right),
\eea
where
\bea
  D&=&1+e^{\eta_1+\eta_1^{*}+R_1}+e^{\eta_1+\eta_2^{*}+\delta_0}+e^{\eta_2+\eta_1^{*}+\delta_0^{*}}+e^{\eta_2+\eta_2^{*}+R_2}+e^{\eta_1+\eta_1^{*}+\eta_2+\eta_2^{*}+R_3},\\
  \eta_j &=& k_j x-(ik_j^2+\omega_j)y+\omega_j t,\quad j=1,2,\\
  e^{\delta_{1j}}&=&\frac{(k_1-k_2)(\alpha_1^{(j)} \kappa_{21}-\alpha_2^{(j)} \kappa_{11})}{(k_1+k_1^*)(k_2+k_1^*)}, \quad
e^{\delta_{2j}}=\frac{(k_2-k_1)(\alpha_2^{(j)} \kappa_{12}-\alpha_1^{(j)} \kappa_{22})}{(k_2+k_2^*)(k_1+k_2^*)}, \\
  e^{Q_{ij}^{(l)}}&=&-{ \frac{(k_i-ib_l)}{(k_j^*+ib_l)}\mu_{ij}}, ~~i,j=1,2, ~~
e^{Q_3^{(l)}}={ \Bigg[\frac{(k_1-ib_l)(k_2-ib_l)}{(k_1^*+ib_l)(k_2^*+ib_l)}\Bigg]e^{R_3}},~ ~~\\
  e^{R_1}&=&\mu_{11},~ e^{R_2}=\mu_{22},~ e^{\delta_0}=\mu_{12},~ e^{\delta_0^*}=\mu_{21},~
e^{R_3}=\frac{|k_1-k_2|^2(\kappa_{11}\kappa_{22}-\kappa_{12}\kappa_{21})}{(k_1+k_1^*)|k_1+k_2^*|^2(k_2+k_2^*)},\label{c8_r3}\\
  \kappa_{ip}&=&{\displaystyle\sum_{j=1}^{m}(\alpha_i^{(j)} \alpha_p^{(j)*})}\bigg(\sum_{l=1}^n\frac{|\rho_l|^2(k_i+k_p^*)}{(k_i-im_l)(k_p^*+im_l)}-(\omega_i+\omega_p^*)\bigg)^{-1},~~\mu_{ip}=\frac{\kappa_{ip}}{(k_i+k_p^*)},\\
  \zeta_l &=&(m_l^2-b_l)t+b_ly+m_lx,~~ i,p=1,2; \;\;j=1,2,...,m;\;\; l=1,2,...,n.
\eea \label{3cmbnd}\ees
We discuss below the two and three short-wave components cases to bring out certain interesting features of the multicomponent LSRI system with $q>2$.

\subsection{Two short-wave components case ($m=1,n=1$)}
For this case, the mixed two-soliton solution with its bright part appearing in the $S^{(1)}$ component and the dark part appearing in the $S^{(2)}$ component can be obtained by putting $m=1$ and $n=1$ in Eq. (\ref{3cmbnd}). The resulting (1b-1d) mixed two-soliton solution is characterized by seven complex parameters ($\alpha_1^{(1)},\alpha_2^{(1)}, k_1, k_2, \omega_1, \omega_2\; \mbox{and}\; \rho_1$) and two real parameters ($b_1$ and $m_1$). This mixed two-soliton solution is restricted by the conditions
\bea
|\rho_1|^2 (k_i+k_p^*) > (\omega_i+\omega_p^*)(k_i-im_1)(k_p^*+im_1), \qquad i,p=1,2,\label{cond1}
\eea
as in the case of the one-soliton solution.

\subsection{Three short-wave components case ($m=2,n=1$)}
\subsubsection{$(2b-1d)$ mixed two-soliton solution}\label{sec2b1d}
\indent First we write down the mixed two-soliton solution with its bright parts in the first two short-wave components $S^{(1)}$ and $S^{(2)}$ while the dark part appears in the third component $S^{(3)}$. For brevity, we mention the straightforward procedure to write down the soliton solution from Eq. (\ref{3cmbnd}) instead of presenting the explicit cumbersome expressions. The explicit forms of bright part of the mixed solitons appearing in $S^{(1)}$ and $S^{(2)}$ can be obtained by putting $j=1$ and $j=2$, respectively, in Eq. (\ref{3cmbnd}a). The dark part of the mixed soliton solution in the $S^{(3)}$ component results from  Eq. (\ref{3cmbnd}b) for $l=1$. In a similar manner, the various quantities appearing in the (2b-1d) mixed two-soliton solution can also be obtained by choosing $m=2$ and $n=1$ in the expressions for $\mu_{ip}$ and $\kappa_{ip}$. The two-soliton solution in this case is characterized by twenty real parameters with the conditions
\bea
|\rho_1|^2 (k_i+k_p^*) > (\omega_i+\omega_p^*) (k_i-im_1)(k_p^*+im_1), \qquad i,p=1,2. \label{cond2}
\eea

\subsubsection{$(1b-2d)$ mixed two-soliton solution}
\indent Another possible split up for the three short-wave components case is to have the bright part of the mixed two-solitons in one component  (say $S^{(1)}$) and  the other two components ($S^{(2)}$ and $S^{(3)}$) comprise of dark parts. The obtained (1b-2d) mixed two-soliton solution can be deduced from Eq. (\ref{3cmbnd}) by putting $m=1$ and $n=2$. This mixed type two-soliton solution is characterized  by sixteen real parameters and is restricted by the conditions
\bea
\displaystyle\sum_{l=1}^{2}\frac{|\rho_l|^2(k_i+k_l^*)}{(k_i-im_l)(k_l^*+im_l)}>(\omega_i+\omega_p^*),\qquad i,p=1,2, \label{cond3}
\eea for obtaining non-singular solutions.\\

We wish to remark that our above analysis can be extended in a straight-forward way to construct  three- as well as multi-soliton solutions. We have indeed obtained the mixed three-soliton solution, but we desist from presenting the solution here due to its cumbersome expression. Also, from the three-soliton solution we identify that the soliton collision is pair-wise and there is no multi-particle effect. Hence a detailed analysis of two-soliton collision is necessary as the higher order soliton interactions can be analyzed in terms of two-soliton collision.

\section{Soliton Interaction}\label{collision}
The multicomponent mixed type two-soliton solution presented in the preceding section contains all the information regarding the dynamics of two solitons in the ($2+1)D$ multicomponent LSRI system. To elucidate the understanding of the collision of mixed solitons, we present the detailed asymptotic analysis of the two and three component cases separately in this section. In particular, we study the interaction of solitons in the $x-y$ plane. A similar approach can also be very well applied to study the collision dynamics in the $x-t$ plane.

We choose the soliton parameters as $k_{1R}>0,~ k_{2R}>0,~ k_{1I}>k_{2I},~ \frac{k_{2R}}{k_{1R}}>\left|\frac{\omega_{2R}}{\omega_{1R}}\right|,~ \frac{k_{2R}k_{2I}}{k_{1R}k_{1I}}>\left|\frac{\omega_{2R}}{\omega_{1R}}\right|$, without loss of generality. For this choice, we find that for a fixed ``$t$" if the soliton, say $s_1$, is localized along the straight line $\eta_{1R}=k_{1R}x+(2 k_{1R}k_{1I}-\omega_{1R})y+\omega_{1R}t \simeq 0$, then $\eta_{2R}$ will tend to $\pm \infty$ as $(x,y) \rightarrow \pm \infty$. Similarly, if the soliton, say $s_2$, is localized along the straight line $\eta_{2R}=k_{2R}x+(2 k_{2R}k_{2I}-\omega_{2R})y+\omega_{2R}t \simeq 0$, then $\eta_{1R} \rightarrow \pm \infty$.
\subsection{Two short-wave components case}
A careful analysis of the asymptotic forms of the mixed two-soliton solution for  the two short-wave components case shows that the intensities of the bright and dark parts of the solitons before and after collision remain unaltered. There occurs only a position-shift in the bright and dark parts of the two colliding solitons. The asymptotic forms of the solitons in these regions are given below.
\\{\it (i) Before Collision ($x,y \rightarrow -\infty$):}
\\\underline{Soliton $s_1$}
\bes\bea
 S_1^{(1)-} &\simeq & A_1^{1-} \mbox{sech}\left(\eta_{1R}+\frac{R_1}{2}\right) e^{i\eta_{1I}},~\\
 S_1^{(2)-} &\simeq & \rho_1 e^{i(\zeta_1+\phi_1^{(1)}+\pi)}\left[\mbox{cos}(\phi_1^{(1)})\mbox{tanh}\left(\eta_{1R}+\frac{R_1}{2}\right)+i~\mbox{sin}(\phi_1^{(1)})\right],\\
  L &\simeq & -2k_{1R}~ \mbox{sech}^2\left(\eta_{1R}+\frac{R_1}{2}\right).
\eea
\underline{Soliton $s_2$}
\bea
 S_2^{(1)-} &\simeq & A_1^{2-} \mbox{sech}\left(\eta_{2R}+\frac{R_3-R_1}{2}\right) e^{i\eta_{2I}},\\
 S_2^{(2)-} &\simeq & \rho_1 e^{i(\zeta_1+\phi_1^{(2)}+2\phi_1^{(1)})}\left[\mbox{cos}(\phi_1^{(2)})~\mbox{tanh}\left(\eta_{2R}+\frac{R_3-R_1}{2}\right)+i~\mbox{sin}(\phi_1^{(2)})\right],\\
  L &\simeq & -2k_{2R}~ \mbox{sech}^2\left(\eta_{2R}+\frac{R_3-R_1}{2}\right).
\eea
{\it (ii) After Collision ($x,y \rightarrow +\infty$):}
\\\underline{Soliton $s_1$}
\bea
  S_1^{(1)+} &\simeq & A_1^{1+} \mbox{sech}\left(\eta_{1R}+\frac{R_3-R_2}{2}\right) e^{i\eta_{1I}},\\
  S_1^{(2)+} &\simeq & \rho_1 e^{i(\zeta_1+\phi_1^{(1)}+2\phi_1^{(2)})}\left[\mbox{cos}(\phi_1^{(1)})~\mbox{tanh}\left(\eta_{1R}+\frac{R_3-R_2}{2}\right)+i~\mbox{sin}(\phi_1^{(1)})\right],\\
  L &\simeq & -2k_{1R} ~\mbox{sech}^2\left(\eta_{1R}+\frac{R_3-R_2}{2}\right).
\eea
\underline{Soliton $s_2$}
\bea
  S_2^{(1)+} &\simeq & A_1^{2+}\mbox{sech}\left(\eta_{2R}+\frac{R_2}{2}\right) e^{i\eta_{2I}},\\
  S_2^{(2)+} &\simeq & \rho_1 e^{i(\zeta_1+\phi_1^{(2)}+\pi)}\left[\mbox{cos}(\phi_1^{(2)})~\mbox{tanh}\left(\eta_{2R}+\frac{R_2}{2}\right)+i~\mbox{sin}(\phi_1^{(2)})\right],\\
  L &\simeq & -2k_{2R}~ \mbox{sech}^2 \left(\eta_{2R}+\frac{R_2}{2}\right),
\eea\label{2casymp}\ees
where $\eta_{j}, \alpha_j^{(1)},~j=1,2$, $R_1$, $R_2$, $R_3$, $\delta_{11}$, and $\delta_{21}$ are defined in Eq. (\ref{3cmbnd}), $\phi_1^{(l)}=\mbox{tan}^{-1}(\frac{k_{lI}-m_1}{k_{lR}})$, $l=1,2$.

The amplitudes of the bright parts of the solitons $s_1$ and $s_2$ before interaction ($A_{1}^{1-},~A_{1}^{2-}$) and their amplitudes after interaction ($A_{1}^{1+},~A_{1}^{2+}$) are given by $A_{1}^{1-}=\frac{\alpha_1^{(1)} e^{-\frac{R_1}{2}}}{2}$, $A_{1}^{2-}=\frac{e^{\delta_{11}-\frac{R_1+R_3}{2}}}{2}$, $A_{1}^{1+}=\frac{e^{\delta_{21}-\frac{R_2+R_3}{2}}}{2}$ and $A_{1}^{2+}=\frac{\alpha_2^{(1)}e^{-\frac{R_2}{2}}}{2}$. Here and in the following, the superscript (subscript) of $A$'s denotes the soliton (component) number and $-$ ($+$) represents the soliton before (after) collision. Upon substitution of the corresponding expressions for $R$'s and $\delta$'s  we find that the intensities of the bright  parts of the solitons $s_1$ and $s_2$ are same before and after collision. Similarly, we find the amplitudes of dark part of mixed solitons appearing in the $S^{(2)}$ component before and after interaction are same and are equal to $\rho_1$. The two colliding solitons $s_1$ and $s_2$ appearing in the bright component ($S^{(1)}$) also experience a position-shift of opposite sense whose magnitude is given by
\bes\bea
|\Phi|&=&\left|\frac{R_3-R_2-R_1}{2}\right|\equiv \left|\ln{\left(\frac{NN^*}{D_1D_1^*}\right)}\right| ,
\eea
\noindent where
\bea
\qquad \qquad N&=&(k_1-k_2)[(k_1-k_2)\rho \rho^*+(\omega_1-\omega_2)(k_1-im_1) (k_2-im_1)],\qquad \qquad \\
  D_1&=&(k_1+k_2^*)[(k_1+k_2^*)\rho \rho^*-(\omega_1+\omega_2^*)(k_1-im_1)(k_2^*+im_1)]
\eea \label{ph2c} \ees
\noindent and $*$ appearing in the superscript represents the complex conjugation. Note that, we require  $k_1\neq k_2$ in the above equations (\ref{ph2c}) for the solitons to undergo collision. Additionally, the dark soliton $s_1$ ($s_2$) experiences a phase-shift $2 \phi_1^{(2)}-\pi$ ($-2 \phi_1^{(1)}+\pi$).

\begin{figure}[h]
\centering\includegraphics[width=0.95\columnwidth]{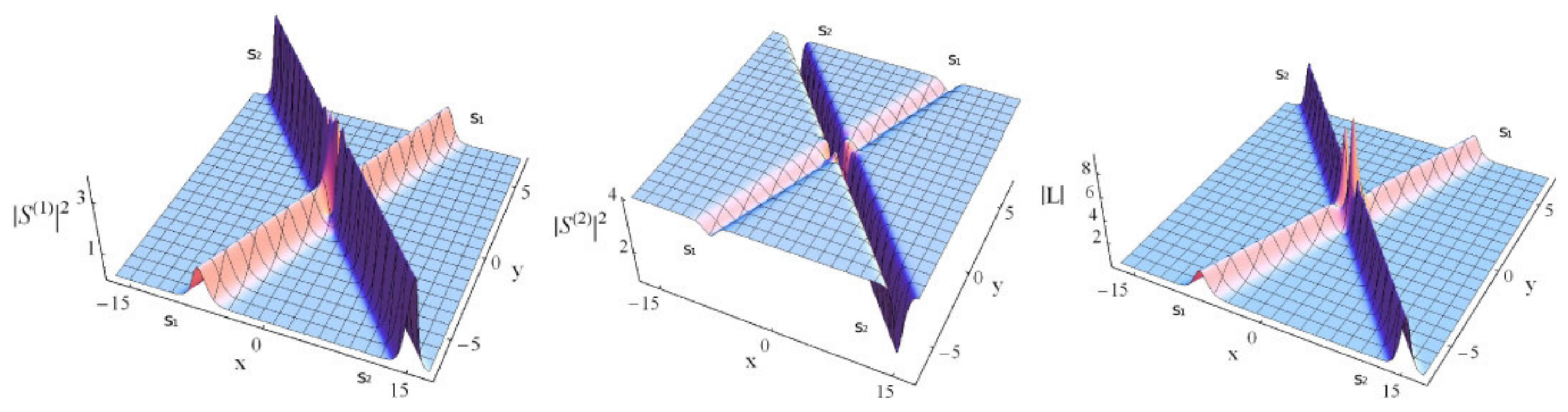}
\caption{Elastic collision of (1b-1d) mixed solitons in two-component LSRI system.}
\label{2celas}
\end{figure}

Thus the mixed solitons in the short-wave components undergo elastic collision. The solitons in the long-wave component also undergo elastic collision with mere position-shift of magnitude $|\Phi|$. The elastic collision of solitons in the two component LSRI system is shown in Fig. \ref{2celas} for the parametric choice $k_1=1-i$, $k_2=1.5+0.75i$, $\omega_1=-1-2i$, $\omega_2=-0.75+i$, $m_1=2$, $\rho_1=2$, $b_1=7$, $\alpha_1^{(1)}=0.3+i$, and $\alpha_2^{(1)}=0.05-i$ at $t=-1$.

We would like to remark that in certain nonlinear integrable dynamical systems ``soliton resonance", that is, two solitons can fuse together after collision or a single soliton can be split up into two solitons, can occur when the shift due to collision of solitons becomes infinity \cite{reso}. In the present two short-wave components case, this corresponds to $|\Phi|\rightarrow \infty$, which is possible for either $|N|^2=0$ or $|D_1|^2=0$. But a careful analysis of the expression (\ref{ph2c}) along with a consideration of the non-singularity condition (\ref{cond1}) into account shows that both  $|N|^2$ and $|D_1|^2$ are positive definite which ensures that $|\Phi|$ is always a finite quantity. Thus it seems that the present solution does not admit the soliton resonance phenomenon. However, it will be  of future interest to look for some special Wronskian type solutions of equation (\ref{model}) and to look for the possibility of such soliton resonance.

\subsection{Three short-wave components case}\label{sec6b}
This case can admit two types of soliton collisions as mentioned in section \ref{twosol} B. The collision scenario depends upon the splitting of mixed solitons into  bright and dark parts among the components and displays interesting dynamical behaviour. To illustrate this, we discuss the soliton collision for this (2b-1d) case in detail.

\subsubsection{(2b-1d) soliton collisions}
\begin{figure}[h]
\centering\includegraphics[width=0.95\columnwidth]{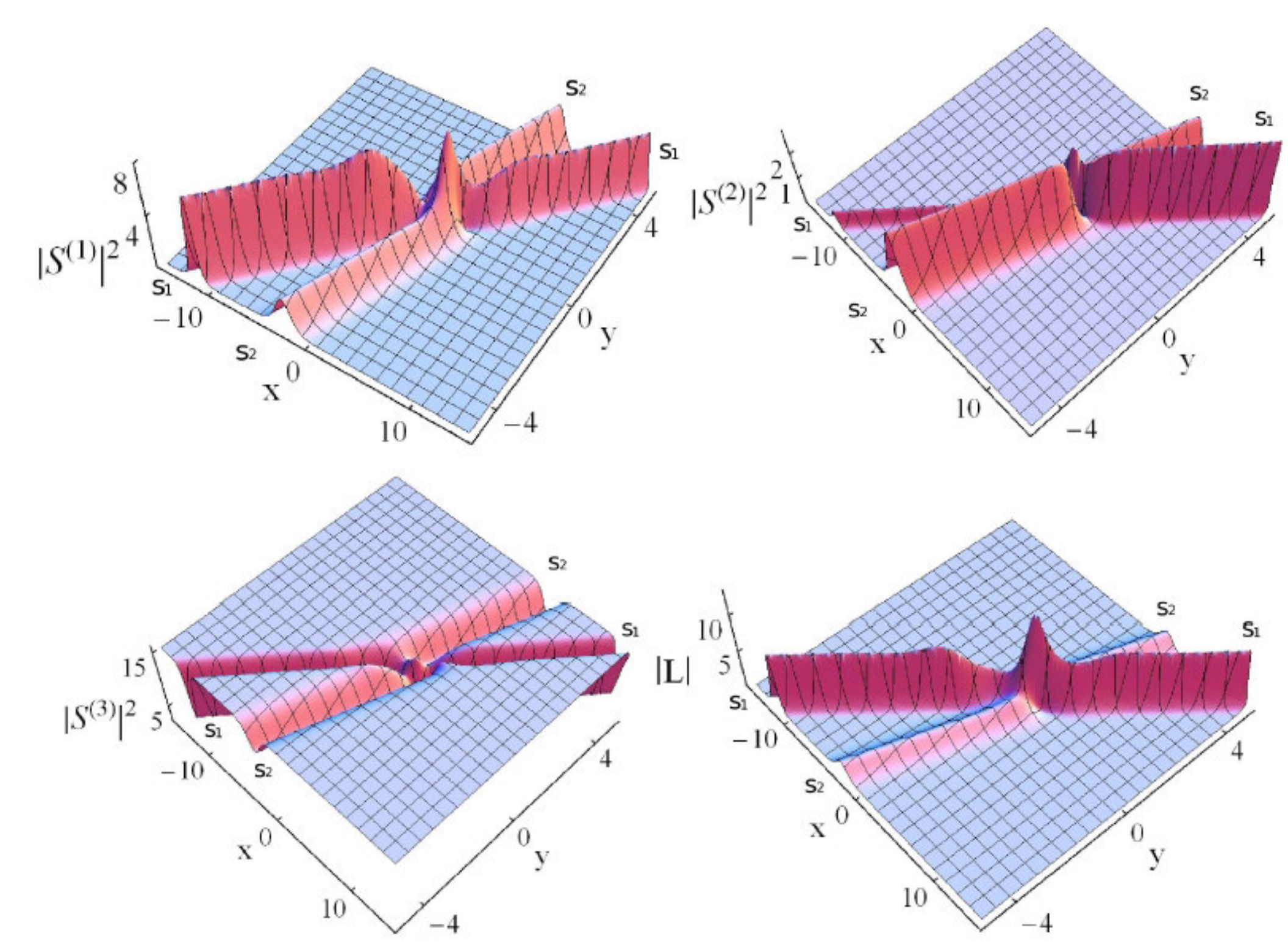}
\caption{Energy exchanging collision of (2b-1d) mixed two-solitons in three-component LSRI system.}
\label{fig2b1d}
\end{figure}

The asymptotic forms of solitons ($s_1$ and $s_2$) before and after collision can be deduced from the exact two-soliton solutions (\ref{3cmbnd}) by putting $m=2$ and $n=1$.
\\{\it (i) Before Collision ($x,y \rightarrow -\infty$):}
\\\underline{Soliton $s_1$}
\bes\bea
   S_1^{(j)-} &\simeq & A_j^{1-} \mbox{sech}\left(\eta_{1R}+\frac{R_1}{2}\right) e^{i\eta_{1I}}, \quad j=1,2,\\
   S_1^{(3)-} &\simeq & \rho_1 e^{i(\zeta_1+\phi_1^{(1)}+\pi)}\left[\mbox{cos}(\phi_1^{(1)})~\mbox{tanh}\left(\eta_{1R}+\frac{R_1}{2}\right)+i~\mbox{sin}(\phi_1^{(1)})\right],\\
  L &\simeq & -2k_{1R}~ \mbox{sech}^2\left(\eta_{1R}+\frac{R_1}{2}\right).
\eea
\underline{Soliton $s_2$}
\bea
  S_2^{(j)-} &\simeq & A_j^{2-} \mbox{sech}\left(\eta_{2R}+\frac{R_3-R_1}{2}\right) e^{i\eta_{2I}}, \quad j=1,2,\\
  S_2^{(3)-} &\simeq & \rho_1 e^{i(\zeta_1+\phi_1^{(2)}+2\phi_1^{(1)})}\left[\mbox{cos}(\phi_1^{(2)})~\mbox{tanh}\left(\eta_{2R}+\frac{R_3-R_1}{2}\right)+i~\mbox{sin}(\phi_1^{(2)})\right],\\
  L &\simeq & -2k_{2R}~ \mbox{sech}^2\left(\eta_{2R}+\frac{R_3-R_1}{2}\right).
\eea
{\it (ii) After Collision ($x,y \rightarrow +\infty$):}
\\\underline{Soliton $s_1$}
\bea
  S_1^{(j)+} &\simeq & A_j^{1+} \mbox{sech}\left(\eta_{1R}+\frac{R_3-R_2}{2}\right) e^{i\eta_{1I}}, \quad j=1,2,\\
  S_1^{(3)+} &\simeq & \rho_1 e^{i(\zeta_1+\phi_1^{(1)}+2\phi_1^{(2)})}\left[\mbox{cos}(\phi_1^{(1)})~\mbox{tanh}\left(\eta_{1R}+\frac{R_3-R_2}{2}\right)+i~\mbox{sin}(\phi_1^{(1)})\right],\\
  L &\simeq & -2k_{1R}~ \mbox{sech}^2\left(\eta_{1R}+\frac{R_3-R_2}{2}\right).
\eea
\noindent\underline{Soliton $s_2$}
\bea
  S_2^{(j)+} &\simeq & A_j^{2+}\mbox{sech}\left(\eta_{2R}+\frac{R_2}{2}\right) e^{i\eta_{2I}}, \quad j=1,2,\\
  S_2^{(3)+} &\simeq & \rho_1 e^{i(\zeta_1+\phi_1^{(2)}+\pi)}\left[\mbox{cos}(\phi_1^{(2)})~\mbox{tanh}\left(\eta_{2R}+\frac{R_2}{2}\right)+i~\mbox{sin}(\phi_1^{(2)})\right],\\
  L &\simeq & -2k_{2R}~ \mbox{sech}^2\left(\eta_{2R}+\frac{R_2}{2}\right).
\eea\ees
The various quantities appearing in the above equations can be obtained from Eq. (\ref{3cmbnd}) by putting $m=2$ and $n=1$. Here, $\phi_1^{(l)}=\mbox{tan}^{-1}(\frac{k_{lI}-m_l}{k_{lR}})$, $l=1,2$. We find that the amplitudes of the bright parts of the two solitons $s_1$ and $s_2$ before and after interaction are related through the relation,
\bes\bea
A_{j}^{l+} &=& T_l^{j} A_{j}^{l-},\qquad  j=1,2, \quad l=1,2,
\eea
where $A_{j}^{1-}=\frac{\alpha_1^{(j)}}{2}e^{-R_1/2}$, $A_{j}^{2-}= \frac{1}{2}e^{\delta_{1j}-\frac{R_1+R_3}{2}}$ and the transition amplitudes $T_l^j$'s are defined as
\bea
T_1^{j} &=& \left(\frac{k_2-k_1}{k_2^*-k_1^*}\right) \left(\frac{k_1^*+k_2}{k_1+k_2^*}\right)^{1/2} \left(\frac{\left(\frac{\alpha_2^{(j)}} {\alpha_1^{(j)}}\right) \lambda_1-1}{\sqrt{1-\lambda_1 \lambda_2}}\right),\\
T_2^{j} &=& \left(\frac{k_1-k_2}{k_1^*-k_2^*}\right) \left(\frac{k_1^*+k_2}{k_1+k_2^*}\right)^{1/2} \left(\frac{\sqrt{1-\lambda_1 \lambda_2}}{\left(\frac{\alpha_1^{(j)}} {\alpha_2^{(j)}}\right) \lambda_2-1 }\right),\;j=1,2,
\eea\ees
in which $\lambda_1=\frac{\kappa_{12}}{\kappa_{22}}$ and $\lambda_2=\frac{\kappa_{21}}{\kappa_{11}}$. Here the $\kappa_{ip}$'s can be obtained from Eq. (\ref{3cmbnd}) by putting $m=2$ and $n=1$. This shows that the intensities of the bright parts of the mixed solitons before and after collision are different in general. The transition amplitudes $T_l^{j}$'s, $l,j=1,2$, become unimodular only for the choice $|{\alpha_1^{(1)}}|/|{\alpha_2^{(1)}}|=|{\alpha_1^{(2)}}|/|{\alpha_2^{(2)}}|$. On the other hand, the intensities of the dark parts of the two solitons $s_1$ and $s_2$  appearing in the third component remain unaltered after collision. Both bright and dark parts of the mixed solitons $s_1$ and $s_2$ experience a position-shift of same magnitude but of different signs. The position-shift experienced by soliton $s_1$ (and $s_2$) is $\Phi_1=\frac{R_3-R_2-R_1}{2}$ (and $\Phi_2=-\Phi_1$), where $R_1,~R_2$ and $R_3$ are defined in Eq. (\ref{3cmbnd}) and the dark solitons $s_1$ and $s_2$ experience phase-shifts $2\phi_1^{(2)}-\pi$ and $-2\phi_1^{(1)}+\pi$, respectively. This shows that the bright components exhibit energy exchanging collisions characterized by an intensity redistribution (energy sharing) among the bright parts of the mixed solitons appearing in first two components and an amplitude dependent position-shift, whereas the dark parts of the two solitons undergo mere elastic collision accompanied by the same position-shift as that of bright  parts. Such a collision scenario is depicted in Fig. \ref{fig2b1d} for the parametric choice $k_1=2-i$, $k_2=1.5-i$, $\omega_1=1+i$, $\omega_2=-1+i$, $m_1=0.7$, $\rho_1=4$, $b_1=2$, $\alpha_1^{(1)}=1.2+0.2i$, $\alpha_2^{(1)}=-1+2i$, $\alpha_1^{(2)}=0.25+0.25i$, and $\alpha_2^{(2)}=-1+i$ at $t=0$. In Fig. \ref{fig2b1d}, the intensity of soliton $s_1$ is suppressed (enhanced) in the $S^{(1)}$ ($S^{(2)}$) component and the reverse occurs for soliton $s_2$. But the solitons appearing in $S^{(3)}$ and $L$ components undergo elastic collision only.

\begin{figure}[h]
\centering\includegraphics[width=0.9\columnwidth]{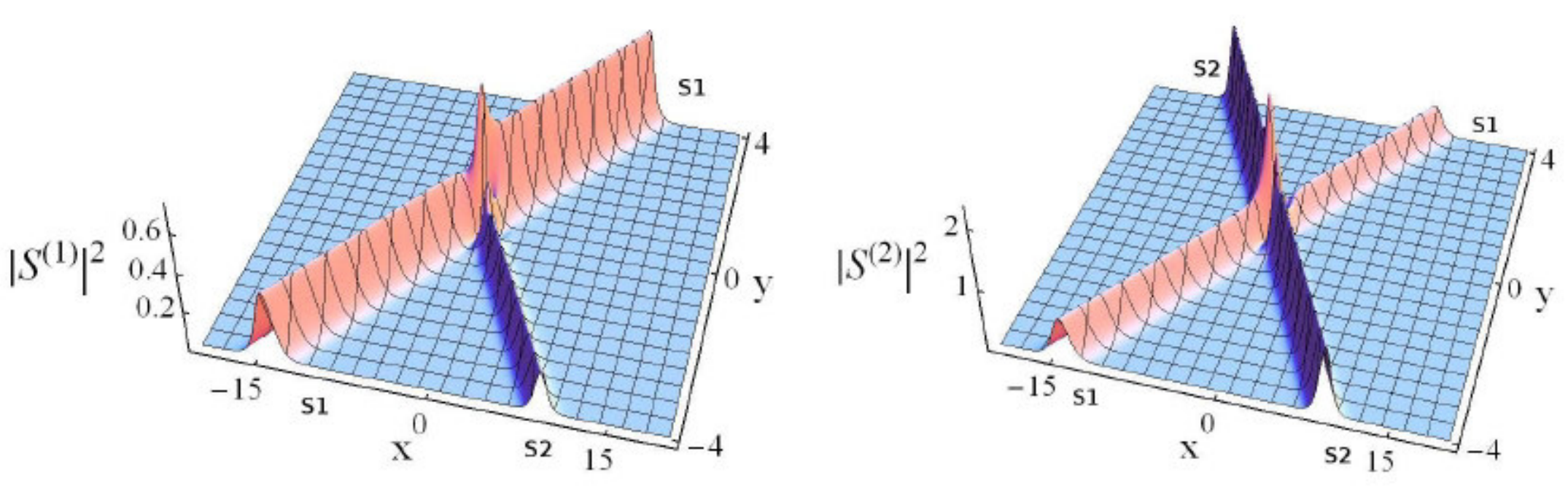}
\caption{Energy sharing collision with complete suppression of soliton $s_2$ after collision in $S^{(1)}$ component of (2b-1d) mixed two-solitons in 3 short-wave LSRI system.}
\label{suppr}
\end{figure}

By performing an analysis similar to that of two short-wave components case here also we could not identify any  resonant interaction.  Additionally, we point out an interesting energy sharing collision with complete suppression of intensity of a particular soliton after collision completely as demonstrated for soliton $s_2$ in a particular component (say $S^{(1)}$) in Fig. \ref{suppr}, with commensurate changes in soliton $s_1$ as well as for the solitons in the other short-wave component $S^{(2)}$. The parameters are chosen as  $k_1=1-2i$, $k_2=1.5+i$, $\omega_1=-1-i$, $\omega_2=-0.75+i$, $m_1=0.7$, $\rho_1=0.5$, $b_1=2$, $\alpha_1^{(1)}=0.5$, $\alpha_2^{(1)}=0.02$, $\alpha_1^{(2)}=0.7$, and $\alpha_2^{(2)}=1$ at $t=-3$. We do not present the collision of dark solitons as it is a standard elastic collision. Note that it is also possible to completely suppress the intensity of a particular soliton in a given component before interaction and can have two solitons after interaction.  However, the physical mechanism behind such fascinating collision is different from that of standard soliton resonance, as pointed out before.
 In fact, this is  due to an intensity redistribution among the components accompanied by finite amplitude dependent position-shift.

\subsubsection{($1b-2d$) soliton collisions}
\begin{figure}[h]
\centering\includegraphics[width=0.9\columnwidth]{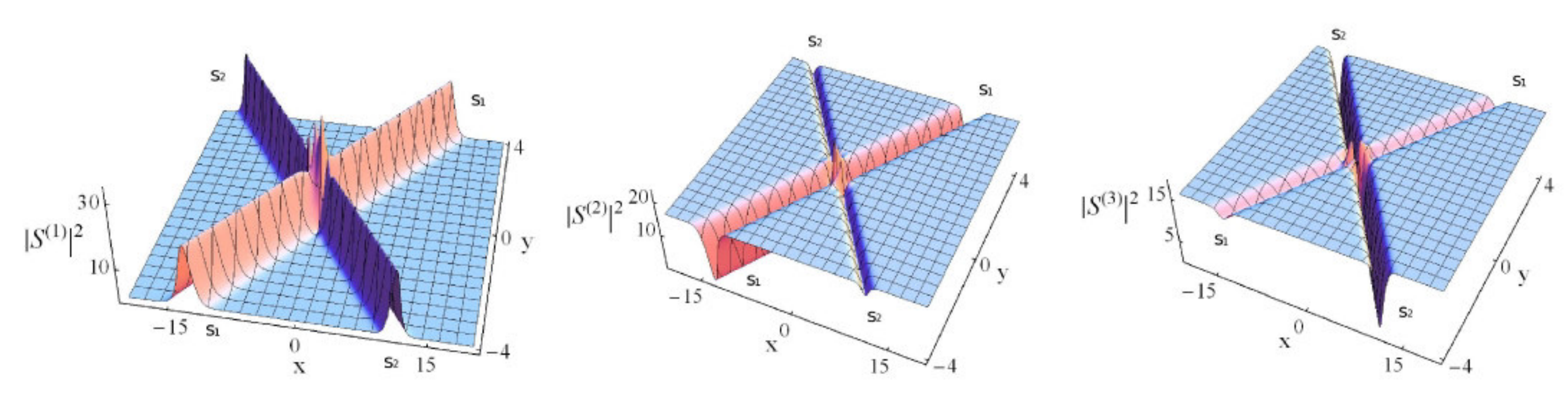}
\caption{Elastic collision of (1b-2d) mixed two-solitons in three-component LSRI system.}
\label{fig1b2delas}
\end{figure}

Next we consider the collision scenario in the three-component LSRI system where the two colliding mixed solitons (say $s_1$ and $s_2$) are comprised of one bright part and two dark parts and appear respectively in the $S^{(1)}$ and ($S^{(2)},S^{(3)}$) components. A careful asymptotic analysis of Eq. (\ref{3cmbnd}) with  $m=1$ and $n=2$ shows that the amplitudes of the bright parts of the mixed solitons before and after collision are given by
\bea
&& \left(A_{1}^{1-},~A_{1}^{2-},~A_{1}^{1+},~A_{1}^{2+}\right) = \left(\frac{\alpha_1^{(1)}}{2}e^{-\frac{R_1}{2}}, ~\frac{1}{2}e^{\delta_{11}-(\frac{R_1+R_3}{2})},~\frac{1}{2}e^{\delta_{21}-(\frac{R_2+R_3}{2})}, ~\frac{\alpha_2^{(1)}}{2}e^{-\frac{R_2}{2}}\right).
\label{12swamps}\eea
Substitution of the expressions for various quantities from Eq. (\ref{3cmbnd}) with $m=1$ in Eq. (\ref{12swamps}) shows that the intensities of the bright parts of the two mixed solitons are same before and after interaction, i.e., $|A_{1}^{j+}|^2=|A_{1}^{j-}|^2,\;j=1,2$. Similarly, the amplitudes of the dark parts of the two mixed solitons before and after collision in the $S^{(2)}$ ($S^{(3)}$) component are same and are equal to $\rho_1$ ($\rho_2$). This clearly indicates that the intensities of the dark parts of the colliding mixed solitons are unaltered during collision. Thus for the (1b-2d) case from the above expressions we observe that both the bright and dark parts of the mixed solitons undergo standard elastic collision of solitons accompanied by position-shifts of magnitude $\left|{(R_3-R_2-R_1)}/{2}\right|$, where $R_1$, $R_2$ and $R_3$ are defined in equation(\ref{3cmbnd}). The phase-shifts of the dark solitons $s_1$ and $s_2$ are $2\phi_1^{(2)}-\pi$ and $-2\phi_1^{(1)}+\pi$, respectively. Such an elastic collision behaviour is shown in Fig. \ref{fig1b2delas} for $k_1=1-2i$, $k_2=1.5+i$, $\omega_1=-1-i$, $\omega_2=-2+i$, $m_1=-2$, $m_2=0.5$, $\rho_1=4$, $\rho_2=4$, $b_1=1$, $b_2=2$, $\alpha_1^{(1)}=0.5-i$, and $\alpha_2^{(1)}=1.4+i$ at $t=-3$. We do not present the plot for the long-wave component in Fig. \ref{fig1b2delas}  as it exhibits the collision process same as that of the $S^{(1)}$ component except for different amplitudes.

Our analysis reveals the interesting fact that the energy exchanging collision of mixed solitons can be realized only in the bright parts  of the mixed solitons in three-component LSRI system and is possible only if the bright parts of the mixed solitons appear at least in two components.

The above analysis can be extended straightforwardly to arbitrary $N$ short-wave components case where the bright parts of the mixed soliton appear in $m$-components and the remaining $(N-m)$ components admit dark parts. It can be shown that the shape-changing (energy exchanging) collision  is possible only if the bright parts of the mixed solitons  appear at least in two short-wave components, i.e. $m\geq2$.

\section{Soliton bound states}\label{secbs}
Soliton bound states are another interesting class of multisoliton solutions. Soliton bound states can be viewed as composite solitons moving with a common speed. Two-soliton bound states in the $(2+1)D$ two component LSRI system (\ref{model}) with $q=2$ can be obtained from Eq. (\ref{3cmbnd}) for the choice $\frac{\omega_{1R}}{k_{1R}}=\frac{\omega_{2R}}{k_{2R}}$ and $k_{1I}=k_{2I}$, $m=1$, $n=1$, and the corresponding solution reads as
\bes\bea
  S^{(1)}=&&\frac{1}{D_1} \left(e^{\frac{l_1+\delta_{21}}{2}} \cosh(\hat{\eta}_{2R}+i\hat{\delta}_{21})+e^{\frac{l_2+\delta_{11}}{2}} \cosh(\hat{\eta}_{1R}+i\hat{\delta}_{11})\right)e^{i\eta_{1I}},\\
  S^{(2)}=&&\frac{\rho_1 e^{i\zeta_1}}{D_1} \left[e^{\frac{Q_3^{(1)}}{2}} \cosh\left(N_1-i\frac{Q_{3I}^{(1)}}{2}\right) + e^{\frac{Q_{11}^{(1)}+Q_{22}^{(1)}}{2}} \cosh(N_2+i\hat{Q}_{11}) \right.\nonumber \\
  &&\left. \qquad \qquad+ e^{\frac{Q_{12}^{(1)}+Q_{21}^{(1)}}{2}} \cosh(N_3+i\hat{Q}_{12}) \right],\\
  D_1=&& e^{\frac{R_3}{2}} \mbox{cosh}\left(\eta_{1R}+\eta_{2R}+\frac{R_3}{2}\right) + e^{\frac{R_1+R_2}{2}} \mbox{cosh}\left(\eta_{1R}-\eta_{2R}+\frac{R_1-R_2}{2}\right) + e^{\frac{\delta_0+\delta_0^*}{2}} \mbox{cos}(\delta_{0I}),\nonumber
\eea\label{bound}\ees
where $\eta_{1R}=k_{1R}\left[x+(2k_{1I}-\frac{\omega_{1R}}{k_{1R}})y+\frac{\omega_{1R}}{k_{1R}}t\right], \;$  $\eta_{2R}=\eta_{1R}(\frac{k_{2R}}{k_{1R}})$, $e^{l_1}=\alpha_1^{(1)}$, $e^{l_2}=\alpha_2^{(1)}$, $\hat{\delta}_{21}=\frac{\delta_{21I}-l_{1I}}{2}$, $\hat{\delta}_{11}=\frac{\delta_{11I}-l_{2I}}{2}$, $\hat{Q}_{11}=\frac{Q_{11I}^{(1)}-Q_{22I}^{(1)}}{2}$, $\hat{Q}_{12}=\frac{Q_{12I}^{(1)}-Q_{21I}^{(1)}}{2}$, $\hat{\eta}_{1R}=\eta_{1R}+\frac{\delta_{11R}-l_{2R}}{2}$, $\hat{\eta}_{2R}=\eta_{2R}+\frac{\delta_{21R}-l_{1R}}{2}$, $N_1=\eta_{1R}+\eta_{2R}+\frac{Q_{3R}^{(1)}}{2}$,  $N_2=\eta_{1R}-\eta_{2R}+\frac{Q_{11R}^{(1)}-Q_{22R}^{(1)}}{2}$, $N_3=\eta_{1I}-\eta_{2I}+\frac{Q_{12I}^{(1)}-Q_{21I}^{(1)}}{2}$, $\eta_{jI}= k_{jI} x- (k_{jR}^2-k_{jI}^2+\omega_{jI})y+\omega_{jI}t$, $j=1,2$,  and all the other quantities found in the above expressions can be deduced from the corresponding quantities appearing in Eq. (\ref{3cmbnd}) by putting $m=1$ and $n=1$. The suffixes $R$ and $I$ appearing in the various quantities in Eq. (\ref{bound}) denote real and imaginary parts, respectively.

\begin{figure}[h]
\centering\includegraphics[width=0.9\columnwidth]{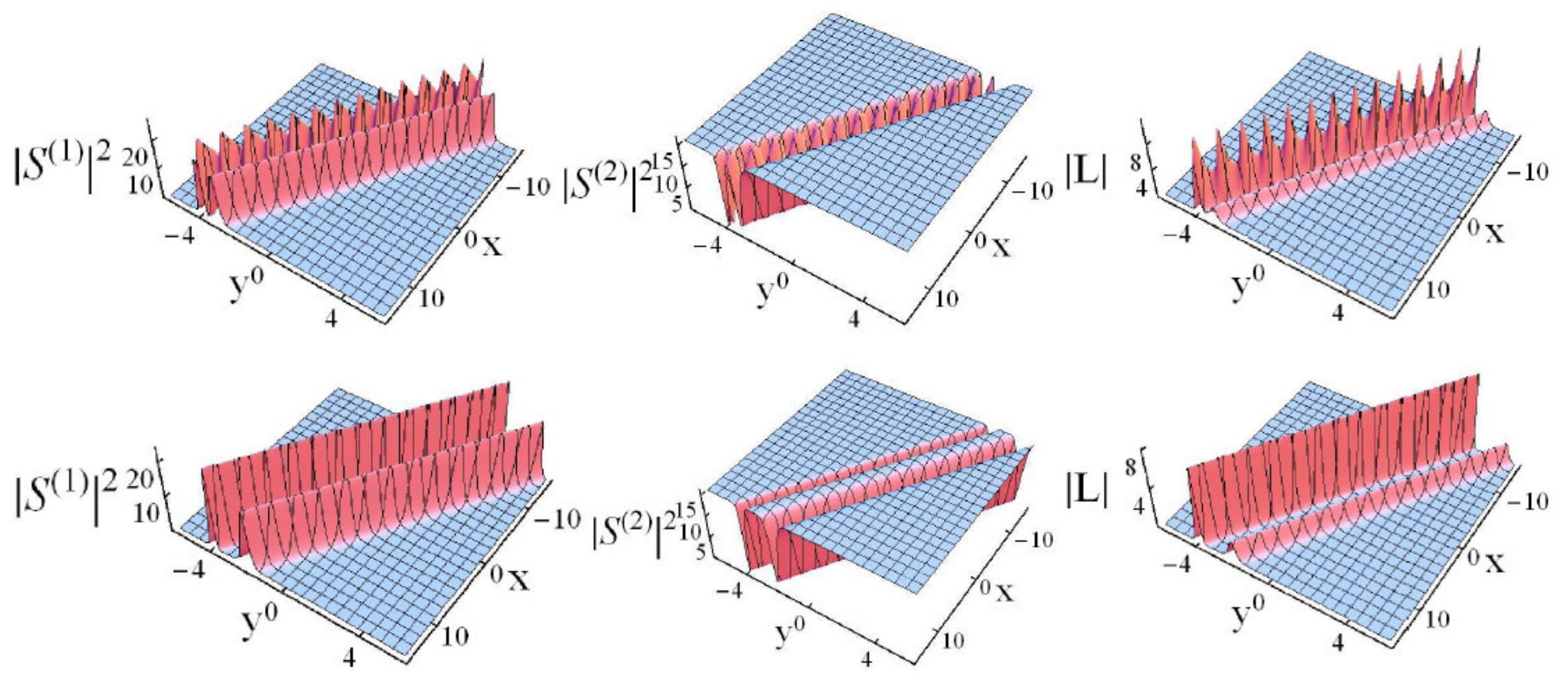}
\caption{Mixed two-soliton bound states in the two-component LSRI system. Top panel: with beating effects, Bottom panel: suppression of beating by tuning the $\alpha_1^{(j)}$-parameters.}
\label{fig9}
\end{figure}

The two-soliton bound state is shown in Fig. \ref{fig9} (top panel) for the choice $k_1=1+i$, $k_2=2+i$, $\omega_1=-2-i$, $\omega_2=-4+7i$, $m_1=0.7,~t=-1$, $\alpha_1^{(1)}=1+i$, $\alpha_2^{(1)}=0.5+i$,  $\rho_1=4$ and $b_1=2$. The bound state solitons display beating effects due to the oscillatory terms in Eq. (\ref{bound}). The beating effects can be suppressed completely by tuning the $\alpha_1^{(j)}$-parameters. This is shown in the bottom panel of Fig. \ref{fig9}, where the parameters are chosen  same as that of the plots in the top panel except for $\alpha_1^{(1)}$, which is now fixed as 0.02. It should be noticed that though the $\alpha$-parameters do not have any observable effects on the single soliton propagation in the two short-wave components case as discussed in section \ref{sec2c1b1d}, they can display significant effects while considering bound state soliton propagation.

\begin{figure}[h]
\centering\includegraphics[width=0.9\columnwidth]{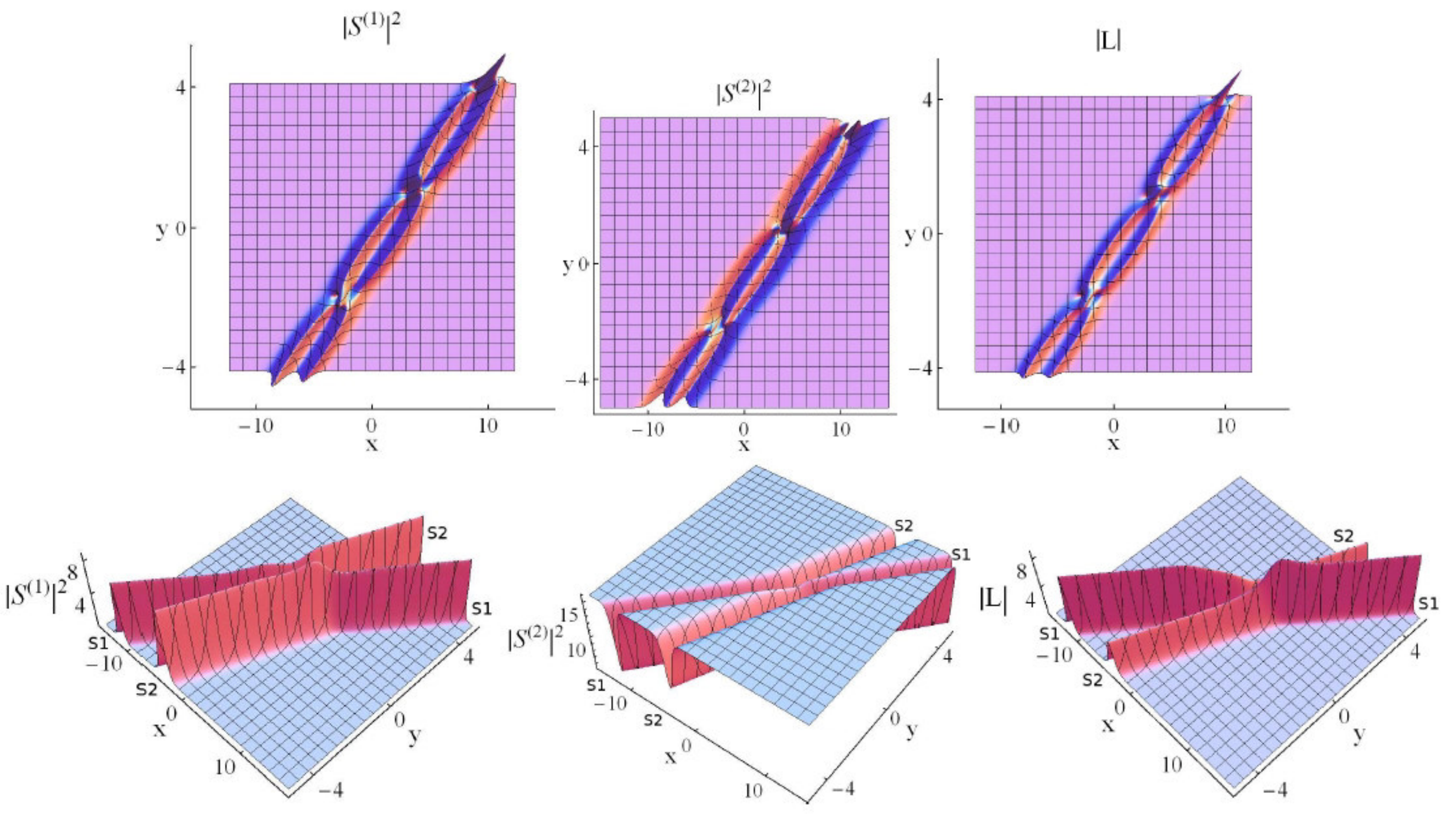}
\caption{Mixed two-soliton bound state for $\omega_1 =\omega_2 = 0$ (top panel) and their transition to colliding solitons for $\omega_1=1+i$, $\omega_2=-1+i$ (bottom panel) in two-component LSRI system. Other parameters are given in the text.}
\label{fig11}
\end{figure}

Now it is of interest to investigate the influence of the $\omega_j,~j=1,2,$ parameters on the bound state soliton dynamics which are arising due to the  higher dimensional nature of the system. In this connection, we assume $\omega_1=\omega_2=0$  and we choose  $k_{1I}=k_{2I}$. This will result in a bound state in the $x-y$ plane which is stationary in time. Such a two-soliton bound state with breathing oscillations in the $x-y$ plane is shown in the top panel of Fig. \ref{fig11} for $\omega_1=\omega_2=0$, $k_1=2-i$, $k_2=1.5-i$, $m_1=0.7$, $\rho_1=4$, $b_1=2$,  $\alpha_1^{(1)}=1.2+0.2i$, and $\alpha_2^{(1)}=1+2i$ at $t=0$. But when $\omega_1$ and $\omega_2$ become different and non-zero, they make the two solitons to undergo collision in the $x-y$ plane. Thus due to the presence of the $\omega_j$-parameters and the higher dimensionality of the system there occurs a transition from bound states to interacting solitons which is shown in the bottom panel of Fig. \ref{fig11}. This also shows that the presence of $\omega_j$ parameters results in a wide range of parameters for which the soliton collision can take place.

\begin{figure}[h]
\centering\includegraphics[width=0.7\columnwidth]{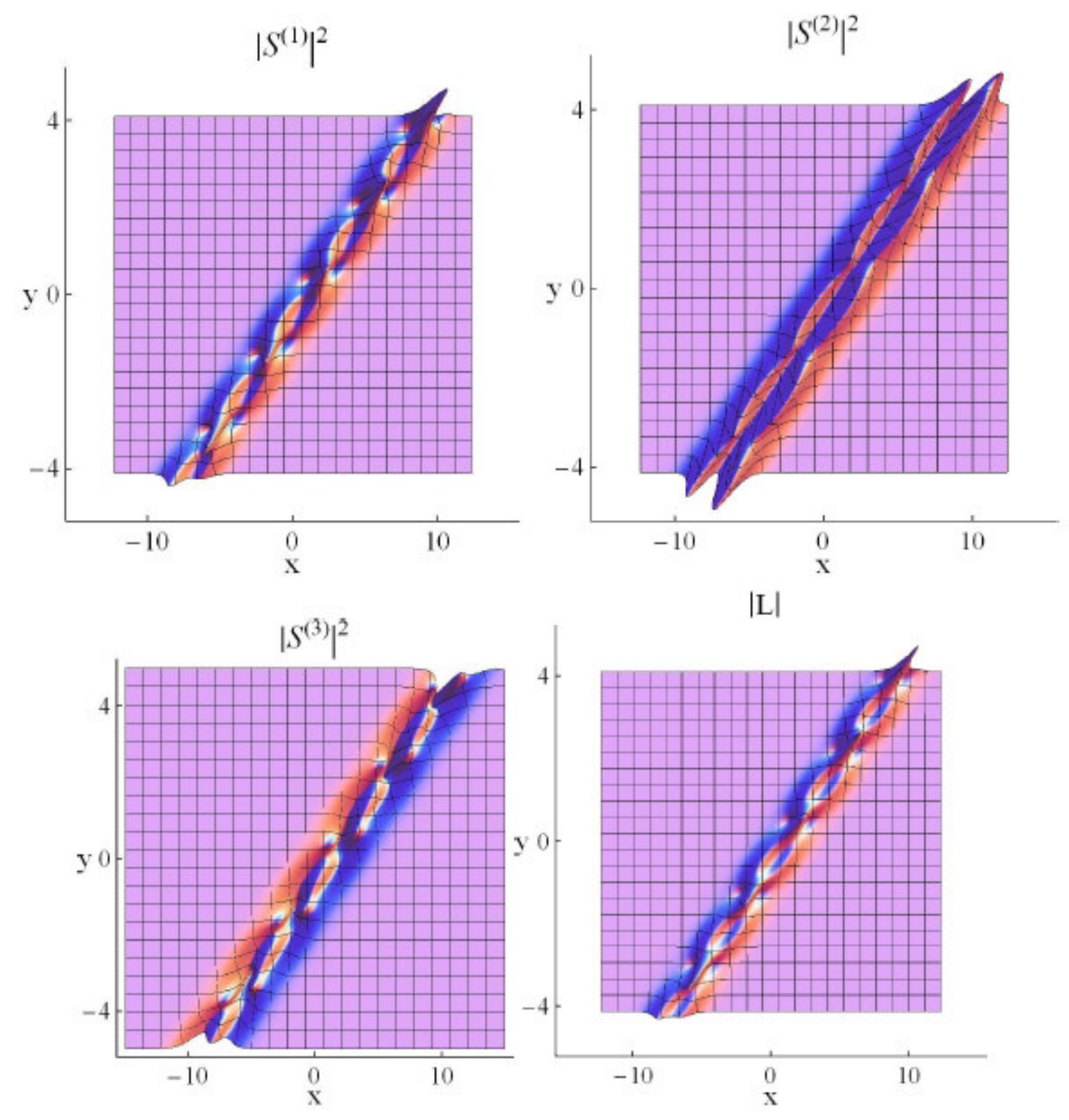}
\caption{(2b-1d) mixed two-soliton bound states in three-component LSRI system for $\omega_1=\omega_2=0$ and their transition to colliding solitons for $\omega_j\neq 0$ is given in Fig. \ref{fig2b1d}.}
\label{fig12}
\end{figure}

The dramatic change in the nature of soliton propagation in the $x-y$ plane due to the presence of $\omega_j$-parameters resulting from the higher dimensional nature of the system (\ref{model}) can exhibit additional features if we consider three short-wave components case. Soliton bound states in $x-y$ plane for $\omega_1=\omega_2=0$, $k_1=2-i$, $k_2=1.5-i$, $m_1=0.7$, $\rho_1=4$, $b_1=2$, $\alpha_1^{(1)}=1.2+0.2i$, $\alpha_2^{(1)}=-1+2i$, $\alpha_1^{(2)}=0.25+0.25i$, and $\alpha_2^{(2)}=-1+i$ at $t=0$ is shown in Fig. \ref{fig12}. For the same parameters, with non-zero $\omega_j$ values, there occurs collision of solitons and this is depicted in Fig. \ref{fig2b1d} for $\omega_1=1+i$ and $\omega_2=-1+i$. Interestingly, the presence of second short-wave component now induces the fascinating collision involving energy exchange among the solitons in short-wave components. Also, the $\alpha$-parameters can be tuned appropriately to suppress the beating effects of the bound soliton states.

\section{Conclusion}
In this paper, we have derived the multicomponent LSRI system for the propagation of weak nonlinear dispersive waves in $(2+1)D$. Then we consider the integrable multicomponent LSRI system in (2+1)-dimensions and obtain mixed (bright-dark) one- and two-soliton solutions using Hirota's method. Our study shows that the bright and dark parts of the mixed solitons in the 2 short-wave components behave like scalar bright and dark solitons. But in three short-wave components case different types of  splitting of the mixed soliton  into bright and dark parts are  possible which makes their dynamics interesting. Study on the collision dynamics of the mixed solitons shows that their collision in two short-wave  components case and in the three short-wave components case with two dark parts and one bright part are elastic. However in the three short-wave components case where the two colliding mixed solitons are split up into  two bright parts  and one dark part, the bright parts undergo energy exchanging collision characterized by intensity redistribution (energy sharing) and amplitude dependent position-shift. This collision process is also influenced by the  soliton parameters of the dark part. The dark parts of the mixed solitons undergo only elastic collision.

Finally, we have considered the soliton bound states. To elucidate the understanding, we explicitly presented the two-soliton bound state expression for $m=1$ and $n=1$. Interestingly, we find that the $\alpha$-parameters which do not show any significant effect on the one-soliton propagation display interesting effects on the bound states. Particularly, they can be profitably used in suppressing the beating effects. Another important observation follows from our above study is that the presence of $\omega_j$-parameters can alter significantly the dynamics of solitons. Specifically, in the absence of $\omega_j$-parameters, which can result due to the higher dimensional nature of the system, there occurs bound state solitons in the $x-y$ plane. But when the $\omega_j$'s are brought into picture the solitons exhibit collision behaviour. Physically, this means that due to the presence of $\omega_j$-parameters the attractive force between the bound solitons vanishes and the solitons pass through each other. It has also been shown that for the  three short-wave components and one long-wave component case, for non-zero values of $\omega_j,~j=1,2,$ one can have energy exchanging collision for the bright parts of the mixed solitons in the $x-y$ plane whereas in the absence of $\omega_j$'s there occurs only stationary bound soliton in the $x-y$ plane.

\section*{Acknowledgement}
T.K. acknowledges the support of the Department of Science and Technology, Government of India, in the form of a research project and also thanks the principal and management of Bishop Heber College for constant support and encouragement. T.K. also thanks K. Sakkaravarthi for his assistance in the preparation of the manuscript. M.V. acknowledges the financial support from UGC-Dr. D. S. Kothari post-doctoral fellowship scheme. The work of M.L. is supported by a DST-IRPHA project. M.L. is also supported by a DST Ramanna Fellowship project and a DAE Ramanna Fellowship.

\end{document}